\documentclass[aps,english,preprint,nofootinbib,preprintnumbers]{revtex4-2}
\usepackage{mathrsfs}
\usepackage{hyperref}
\hypersetup{
    colorlinks=true,
    linkcolor=blue,
    filecolor=magenta,      
    urlcolor=cyan,
    pdfpagemode=FullScreen,
    }
\usepackage{graphicx}
\usepackage{amsmath, amssymb,bbm}
\usepackage{babel}
\usepackage{color}
\usepackage{slashed}
\usepackage[T1]{fontenc}
\usepackage{titlesec}
\def\beqn{\begin{eqnarray}}
\def\eeqn{\end{eqnarray}}
\def\barr{\begin{array}}
\def\earr{\end{array}}
\def\btab{\begin{tabular}}
\def\etab{\end{tabular}}
\def\bite{\begin{itemize}}
\def\eite{\end{itemize}}
\def\bcen{\begin{center}}
\def\ecen{\end{center}}

\begin{document}

\title{Hybrid analysis of radiative corrections to neutron decay with current algebra and effective field theory}

\author{Chien-Yeah Seng$^{1,2}$}

\affiliation{$^{1}$Facility for Rare Isotope Beams, Michigan State University, East Lansing, MI 48824, USA}
\affiliation{$^{2}$Department of Physics, University of Washington,
	Seattle, WA 98195-1560, USA}

\date{\today}

\begin{abstract}
	
We introduce a useful framework for high-precision studies of the neutron beta decay by merging the current algebra description and the fixed-order effective field theory calculation of the electroweak radiative corrections to the neutron axial form factor. We discuss the advantages of this hybrid method and show that it only requires a minimal amount of lattice QCD inputs to achieve a $10^{-4}$ theory accuracy for the Standard Model prediction of the neutron lifetime and the axial-to-vector coupling ratio $\lambda$, both important to the search for physics beyond the Standard Model.

\end{abstract}

\maketitle
\newpage


\section{Introduction}

Beta decays of strongly-interacting systems, e.g. pions, free neutron and nuclei, provide a perfect avenue for high-precision tests of the Standard Model (SM) at low energies and searches for physics beyond the Standard Model (BSM). As an example, they allow high-precision extractions of the Cabibbo-Kobayashi-Maskawa (CKM) matrix element $V_{ud}$, which can be used for the test of the CKM unitarity. Decays of spinful particles offer more experimental observables which impose further constraints on BSM physics; for instance, the axial-to-vector coupling ratio $\lambda\equiv g_A/g_V$ in free neutron decay provides a sensitive probe of right-handed currents~\cite{Bhattacharya:2011qm,Alioli:2017ces}. Furthermore, free neutron holds an advantage compared to nuclei for being free from nuclear-structure uncertainties, and first-principles calculations with lattice QCD are more reliable.

Similar to other experiments at the precision frontier, measurements of the neutron decay lifetime $\tau_n$ and $\lambda$, both currently reaching $10^{-4}$ precision~\cite{ParticleDataGroup:2022pth}, must be accompanied by equally-precise calculations of SM theory inputs in order to identify small BSM effects from experimental results. At $10^{-4}$, there are three SM corrections to free neutron decay that require attention: (1) recoil corrections, (2) isospin-symmetry-breaking (ISB) corrections, and (3) radiative corrections (RC), and in this paper we focus on the last, which consists of virtual loop corrections and bremsstrahlung corrections. Historically, it is divided into ``outer'' and ``inner'' corrections~\cite{Wilkinson:1970cdv}; the outer RC modifies the beta decay spectrum and could be computed using elementary Quantum Electrodynamics (QED) assuming point-charge interactions, while the inner RC modifies the vector and axial coupling constants and involves non-perturbative Quantum Chromodynamics (QCD). 
For many years, the study of RC in free neutron and nuclear beta decays has been based on the current algebra formalism by Sirlin~\cite{Sirlin:1974ni} (which we denote as ``Sirlin's representation'', SR), which identified the form factor (FF) corrections and the $\gamma W$-box diagram as the only sources of non-trivial inner RC. Supplemented by recent improved analysis of the vector $\gamma W$-box diagram~\cite{Seng:2018yzq,Seng:2018qru,Seng:2020wjq,Czarnecki:2019mwq,Shiells:2020fqp,Hayen:2020cxh,Ma:2023kfr}, a substantial improvement in the theory precision of the extraction of $V_{ud}$ was achieved. 

Effective field theory (EFT) provides an alternative pathway to the problem; by writing down the most general Lagrangian with the relevant degrees of freedom (DOFs), one may compute the RC-corrected decay amplitude using elementary Feynman rules, with the theory precision controlled by the built-in power-counting scheme of the EFT. 
In the first attempt in Ref.~\cite{Ando:2004rk}, a pionless EFT was adopted, with DOFs being nucleons, leptons and photon. Natural scales in this formalism are $q\sim m_n-m_p\sim 1$~MeV and $m_N\sim 1$~GeV, 
which means the natural expansion parameter of the EFT is $q/m_N\sim 10^{-3}$. With that one could obtain the outer RC as well as the ``kinematic'' recoil corrections. However, the lack of an intermediate scale between $q$ and $m_N$ makes it unable to describe important physics such as nucleon structure corrections induced by pion loops. To incorporate the latter, one needs to start from an EFT that includes pions as dynamical DOFs, namely the chiral perturbation theory (ChPT). This was performed in Ref.\cite{Cirigliano:2022hob} which observed, for the first time, a percent-level inner RC to the axial coupling constant $g_A$ that comes from pion-loop contributions. While not affecting the extraction of $V_{ud}$ from neutron beta decay, it plays an important role when the experimentally-measured $g_A$ is compared to the lattice-calculated, pure-hadronic $\mathring{g}_A$~\cite{FlavourLatticeAveragingGroupFLAG:2021npn} in order to constrain new physics. 

It should be stressed that this unexpectedly large RC was in fact incorporated in SR; it is contained in the so-called ``three-point function'' contribution to the FF correction, but just mistakenly regarded to be small due a misinterpretation of the partially-conserved-axial-current (PCAC) relation. In principle, SR provides an explicit expression of the three-point function in terms of hadronic matrix elements which can be directly computed on lattice; however, in practice this is difficult due to the complexity of such matrix elements. To that end, it is beneficial to adopt a hybrid analysis which combines SR and EFT description~\cite{Seng:2019lxf}. The key idea is to study not the full RC, but only FF RC which is not easy to be computed on lattice, using EFT. First, the current algebra formalism does not rely on a chiral expansion, so the non-FF corrections in this formalism do not suffer from theory uncertainties from such an expansion (although in principle the SU(2) chiral expansion scales in powers of $m_\pi/\Lambda_\chi$, where $\Lambda_\chi\sim 1$~GeV is the chiral symmetry breaking scale, and converges rapidly). Second, SR provides a more convenient basis to connect to lattice QCD as the lepton piece decouples from its master formula, so one may concentrate on pure hadronic matrix elements without introducing dynamical leptons. This strategy was proven successful in semileptonic kaon decay, where a reduction of the theory uncertainty of long-distance RC by almost an order of magnitude was achieved~\cite{Seng:2020jtz,Ma:2021azh,Seng:2021boy,Seng:2021wcf,Seng:2022wcw}. 

In this work, we apply the aforementioned hybrid analysis to the decay of free neutron. While the outer RC and the inner RC to the decay lifetime are not affected, its provides further insights to the percent-level inner RC to $g_A$ discovered in Ref.\cite{Cirigliano:2022hob}. More specifically, we prove our assertion in the paragraph above, that this large correction is contained entirely in the ``three-point function'' in SR that fully decouples with the lepton piece in the decay process. To that end, we point out that if the matrix element $\langle n|(J_W^\mu)_A|n\rangle_{\text{QCD+QED}}$, with $(J_W^\mu)_A$ the axial charged weak current, can be computed on lattice (with a small photon mass $m_\gamma$ as an infrared (IR)-regulator), then one can fully pin down this large correction since all other QED corrections to the matrix element are analytically calculable (to the order $10^{-4}$) in SR. With that one avoids computing a more complicated five-point correlation function on lattice, and makes a realistic comparison between $\mathring{g}_A$ and $g_A$ at sub-percent level precision feasible in the foreseeable future.

The content of this work is arranged as follows.
In Sec.\ref{sec:RCinSR} we introduce the general structure of SR, identify the terms that require inputs from non-perturbative QCD, and define the ``outer'' and ``inner'' corrections rigorously. In Secs.\ref{sec:residual}, \ref{sec:EFT}, \ref{sec:HBChPT} we introduce the EFT formalism based on Heavy Baryon Chiral Perturbation Theory (HBChPT), and use it to compute the RC to the axial FF, which provides useful information on the inner correction to $g_A$. In Sec.\ref{sec:LECs} we compare the expressions of the hybrid analysis to the pure-EFT calculation, which allows us to pin down a subset of low energy constants (LECs) in the latter; we also discuss the future lattice QCD input needed to determine the remaining LECs relevant for the neutron beta decay. We draw our conclusions in Sec.\ref{sec:summary}.

\section{Radiative corrections in Sirlin's representation\label{sec:RCinSR}}

In this section we introduce the basic theory framework for the free neutron decay and its RC.
We start by writing down the tree-level decay amplitude: 
\begin{equation}
\mathfrak{M}_0=-\frac{G_FV_{ud}}{\sqrt{2}}L_\lambda F^\lambda~,
\end{equation}
with $G_F=1.1663787(6)\times 10^{-5}$~GeV$^{-2}$ the Fermi constant obtained from muon decay~\cite{MuLan:2012sih}, and $L_\lambda=\bar{u}_e\gamma_\lambda(1-\gamma_5)v_\nu$ the leptonic charged weak current. The single-nucleon matrix element of the hadronic charged weak current $J_W^\mu=\bar{u}\gamma^\mu(1-\gamma_5)d$ consists of four FFs: vector, axial, weak magnetism and pseudoscalar (excluding second-class currents). Upon neglecting recoil effects, the last two form factors can be dropped while the vector and axial form factors can be approximated by their values at zero momentum exchange. This gives:
\begin{equation}
F^\mu\equiv\langle p|J_W^\mu(0)|n\rangle \approx\bar{u}(p_p)\gamma^\mu\left[\mathring{g}_V+\mathring{g}_A\gamma_5\right]u(p_n)\equiv F_V^\mu+F_A^\mu~,\label{eq:CCME}
\end{equation}
where $\mathring{g}_V$ and $\mathring{g}_A$ are the pure-QCD vector and axial coupling constants. In this paper we adopt the experimentalists' sign convention, $\mathring{g}_A<0$. 

\subsection{$\mathcal{O}(\alpha)$ radiative corrections}

\begin{figure}[t]
	\begin{center}
		\includegraphics[width=0.7\columnwidth]{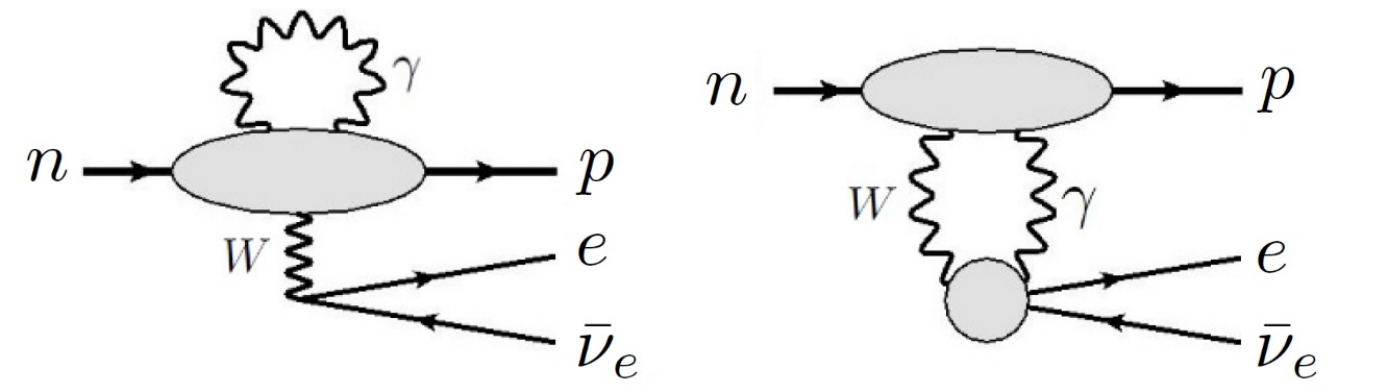}
		\caption{Non-trivial one-loop RCs to neutron beta decay. Left: Form factor corrections, right: $\gamma W$-box diagram.}
		\label{fig:loops}
	\end{center}
\end{figure}

We now discuss the $\mathcal
O(\alpha)$ SM electroweak virtual RC to the decay amplitude based on SR. While details can be found in Ref.~\cite{Seng:2021syx}, here we just summarize the most important results. One can show that, the only pair of one-loop diagrams that are not analytically (exactly or perturbatively) calculable are the electromagnetic RC to the weak FFs and the $\gamma W$-box diagram, depicted by the first and second diagram in Fig.\ref{fig:loops} respectively. 
The RC-corrected $n\rightarrow pe\bar{\nu}_e$ decay amplitude in SR reads:
\begin{eqnarray}
\mathfrak{M}&=&\sqrt{1+\Delta_R^U}\mathfrak{M}_0+\frac{\alpha}{4\pi}\left[3\ln\frac{m_N}{m_e}+2\ln\frac{m_e}{m_\gamma}-\frac{3}{4}\right]\mathfrak{M}_0+\left(\delta \mathfrak{M}_2+\delta \mathfrak{M}_{\gamma W}^a\right)_\mathrm{int}\nonumber\\
&&-\frac{G_FV_{ud}}{\sqrt{2}}\delta F_{3\text{pt}}^\mu L_\mu+\delta \mathfrak{M}_{\gamma W}^b~,\label{eq:Sirlinrep}
\end{eqnarray}
and we explain the notations as follows. First, the quantity
\begin{equation}
\Delta_R^U=\frac{\alpha}{2\pi}\left[3\ln\frac{m_Z}{m_N}+\ln\frac{m_Z}{m_W}+\tilde{a}_g\right]+\delta_\text{HO}^\text{QED}\label{eq:universal}
\end{equation}
represents a universal (U),  analytically-calculable virtual RC proportional to the tree-level amplitude; $\tilde{a}_g\approx -0.083$ is a small perturbative QCD correction, and $\delta_\text{HO}^\text{QED}= 0.00109(10)$ represents a leading higher-order QED effects~\cite{Czarnecki:2004cw} which we include for accuracy\footnote{A more recent analysis suggested a slightly larger value of $\Delta_R^U$, which was partially compensated by a redefinition of the Fermi function and led to a +0.026\% shift of the neutron decay rate~\cite{Cirigliano:2023fnz}}. The second squared bracket in Eq.\eqref{eq:Sirlinrep} collects the other part of the analytically-known RC which will later be grouped into the ``outer correction'' following historical convention. Here we introduce a fictitious photon mass $m_\gamma$ to regularize IR-divergences. 

The remaining terms in $\mathfrak{M}$ are sensitive to physics at and below the hadron scale. To explain their meaning we need to introduce a few more prescriptions. First, using the on-mass-shell perturbation theory~\cite{Brown:1970dd}, one splits the FF correction (i.e. the first diagram in Fig.\ref{fig:loops}) into two terms:
\begin{equation}
\delta F^\mu=\delta F_\text{2pt}^\mu+\delta F_\text{3pt}^\mu~,\label{eq:Fmusplit}
\end{equation}
which we name as the ``two-point function'' and ``three-point function'' respectively. The two-point function takes the following form for neutron:
\begin{equation}
\delta F_{\text{2pt}}^\lambda=\frac{e^2}{2}\int \frac{d^4k}{(2\pi)^4}\frac{\partial}{\partial k_\lambda}\left(\frac{m_W^2}{m_W^2-k^2}\frac{1}{k^2-m_\gamma^2}\right)T^\mu_{\:\:\mu}~,\label{eq:2pt}
\end{equation}
where
\begin{equation}
T^{\mu\nu}\equiv \int d^4x e^{ik\cdot x}\langle p|T[J_\text{em}^\mu(x)J_W^\nu(0)]|n\rangle
\end{equation}
is a ``generalized Compton tensor'' consists of the time-order product of the electromagnetic and charged weak current. 
The definition of the ``three-point function'' is more complicated and consists of two terms:
\begin{equation}
\delta F_\text{3pt}^\lambda=-\lim_{\bar{q}\rightarrow q}i\bar{q}_\nu\frac{\partial}{\partial\bar{q}_\lambda}\left[\bar{T}^\nu-B^\nu\right]+\lim_{\bar{q}\rightarrow  q}i\frac{\partial}{\partial\bar{q}_\lambda}\left[D-\bar{q}\cdot B\right]~,
\label{eq:3pt}
\end{equation}
where $q\equiv p_n-p_p$ is the nucleon momentum exchange, and 
\begin{eqnarray}
\bar{T}^\mu&=&\frac{e^2}{2}\int\frac{d^4k}{(2\pi)^4}\frac{m_W^2}{m_W^2-k^2}\frac{1}{k^2-m_\gamma^2}\int d^4x e^{i\bar{q}\cdot x} d^4y e^{ik\cdot y}\langle p|T\{J_W^\mu(x)J_\text{em}^\nu(y)J_\nu^\text{em}(0)\}|n\rangle\nonumber\\
D&=&\frac{ie^2}{2}\int\frac{d^4k}{(2\pi)^4}\frac{m_W^2}{m_W^2-k^2}\frac{1}{k^2-m_\gamma^2}\int d^4x e^{i\bar{q}\cdot x} d^4y e^{ik\cdot y}\langle p|T\{\partial\cdot J_W(x)J_\text{em}^\nu(y)J_\nu^\text{em}(0)\}|n\rangle\nonumber\\
B^\mu&=&-\bar{u}_p\left[\frac{i\delta m_p}{\slashed{p}_n-\bar{\slashed{q}}-m_p}\mathfrak{T}^\mu+\mathfrak{T}^\mu\frac{i\delta m_n}{\slashed{p}_p+\bar{\slashed{q}}-m_n}\right]u_n~,\label{eq:3ptME}
\end{eqnarray}
with $\delta m_{p,n}$ the nucleon mass shift due to electromagnetic corrections, and $\mathfrak{T}^\mu(\bar{q})$ the single-nucleon weak vertex function that consists of the vector, weak magnetic, axial and pseudoscalar form factors: 
\begin{equation}
	\mathfrak{T}^\mu(\bar{q})=F_1^W(\bar{q}^2)\gamma^\mu-\frac{i}{2m_N}F_2^W(\bar{q}^2)\sigma^{\mu\nu}\bar{q}_\nu+G_A(\bar{q}^2)\gamma^\mu\gamma_5+\frac{G_P(\bar{q}^2)}{2m_N}\gamma_5 \bar{q}^\mu~.
\end{equation}
In particular, the pseudoscalar form factor is related to the axial form factor as:
\begin{equation}
	\frac{G_P(\bar{q}^2)}{2m_N}=\frac{2m_NG_A(\bar{q}^2)}{\bar{q}^2-m_\pi^2}
\end{equation}
to ensure the conservation of the axial current in the chiral limit.
The $B$-terms in Eq.\eqref{eq:3pt} remove the poles in $\bar{T}^\nu$ and $D$ at $\bar{q}\rightarrow q$.    
Finally, we split the $\gamma W$-box diagram, i.e. the second diagram in Fig.\ref{fig:loops}, into two terms:
\begin{equation}
\delta \mathfrak{M}_{\gamma W}=\delta \mathfrak{M}_{\gamma W}^a+\delta \mathfrak{M}_{\gamma W}^b~,
\end{equation}
the (b) term carries an $\epsilon$-tensor from the lepton structure while the (a) term does not.

Now we can discuss the remaining terms in Eq.\eqref{eq:Sirlinrep}. First, a partial cancellation occurs when combining $\delta F_\text{2pt}^\mu$ and $\delta \mathfrak{M}_{\gamma W}^a$, resulting in an analytically-calculable piece (which is already included in the first two terms in Eq.\eqref{eq:Sirlinrep}) and a ``residual integral'' that cannot be simply reduced to something proportional to $\mathfrak{M}_0$:
\begin{eqnarray}
\left(\delta \mathfrak{M}_2+\delta \mathfrak{M}_{\gamma W}^a\right)_\mathrm{int}&=&-\frac{G_FV_{ud}e^2}{\sqrt{2}}L_\lambda\int\frac{d^4k}{(2\pi)^4}\frac{1}{(p_e-k)^2-m_e^2}\frac{1}{k^{ 2}-m_\gamma^2}\nonumber\\
&&\times\left\{\frac{2p_e\cdot kk^{\lambda}}{k^{2}-m_\gamma^2}T^\mu_{\:\:\mu}+2p_{e\mu}T^{\mu\lambda}-(p_n-p_p)_\mu T^{\lambda\mu}+i\Gamma^\lambda\right\}~,\label{eq:resintegral}
\end{eqnarray}
where 
\begin{equation}
\Gamma^{\mu}\equiv \int d^4x e^{ik\cdot x}\langle p|T[J_\text{em}^\mu(x)\partial\cdot J_W(0)]|n\rangle 
\end{equation}
is similar to $T^{\mu\nu}$ except that the charged current is replaced by its total derivative. They satisfy the following Ward identities:
\begin{equation}
k_\mu T^{\mu\nu}=iF^\nu~,~(k-q)_\nu T^{\mu\nu}=iF^\mu-i\Gamma^\mu~.\label{eq:Ward}
\end{equation}
The next non-trivial integral is the part of the $\gamma W$-box diagram with an $\epsilon$-tensor:
\begin{eqnarray}
\delta \mathfrak{M}_{\gamma W}^b&=&-i\frac{G_FV_{ud}e^2}{\sqrt{2}}L_\lambda\int\frac{d^4k}{(2\pi)^4}\frac{m_W^2}{m_W^2-k^{ 2}}\frac{\epsilon^{\mu\nu\alpha\lambda}k_\alpha}{[(p_e-k)^2-m_e^2]k^{2}}T_{\mu\nu}~.\label{eq:gammaWb}
\end{eqnarray}
SR allows us to identify these two integrals, together with the three-point function in Eq.\eqref{eq:3pt}, as the only non-trivial quantities in the $\mathcal{O}(\alpha)$ virtual RC. 

\begin{figure}[t]
	\begin{center}
		\includegraphics[width=0.3\columnwidth]{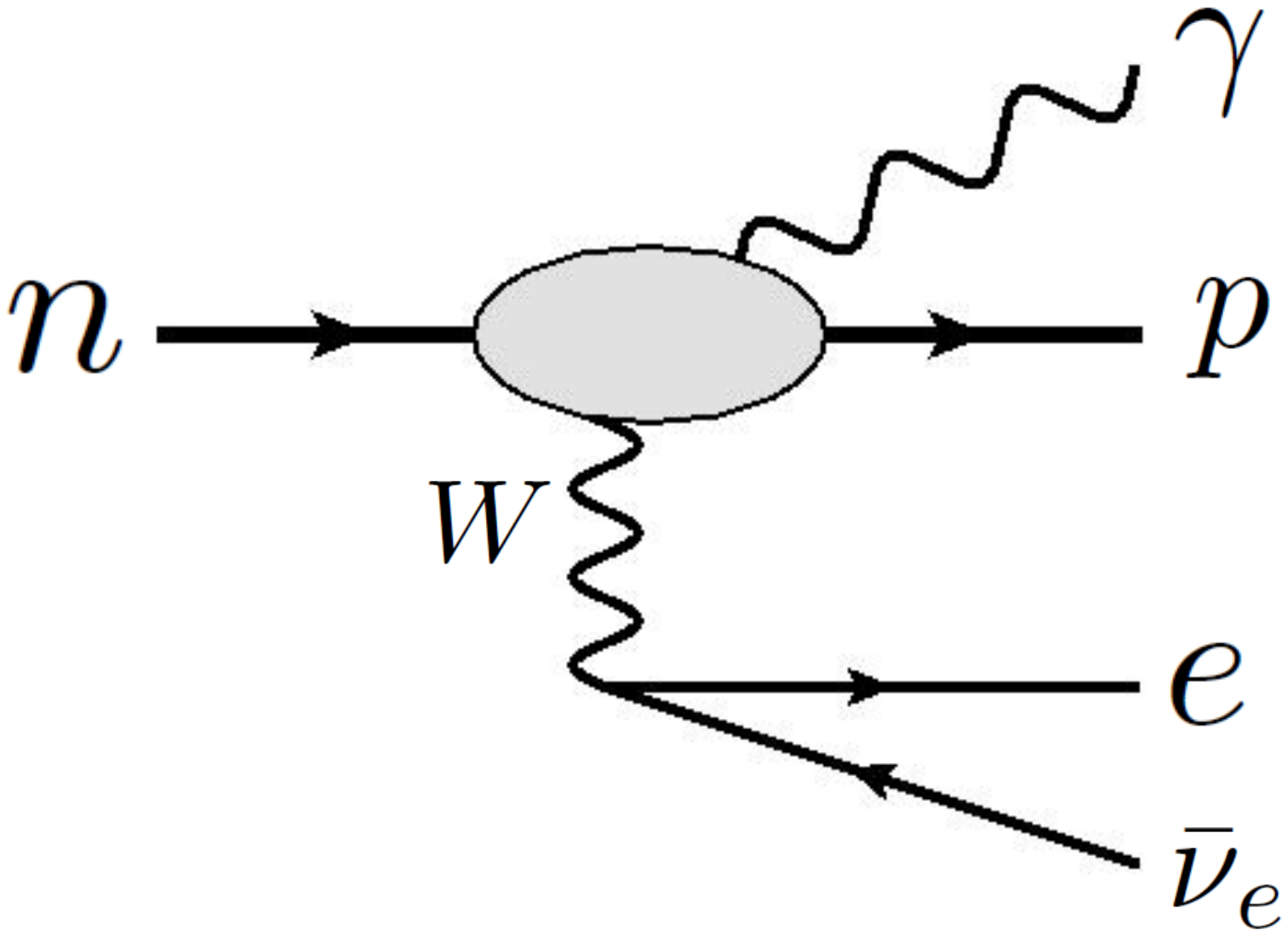}
		\includegraphics[width=0.3\columnwidth]{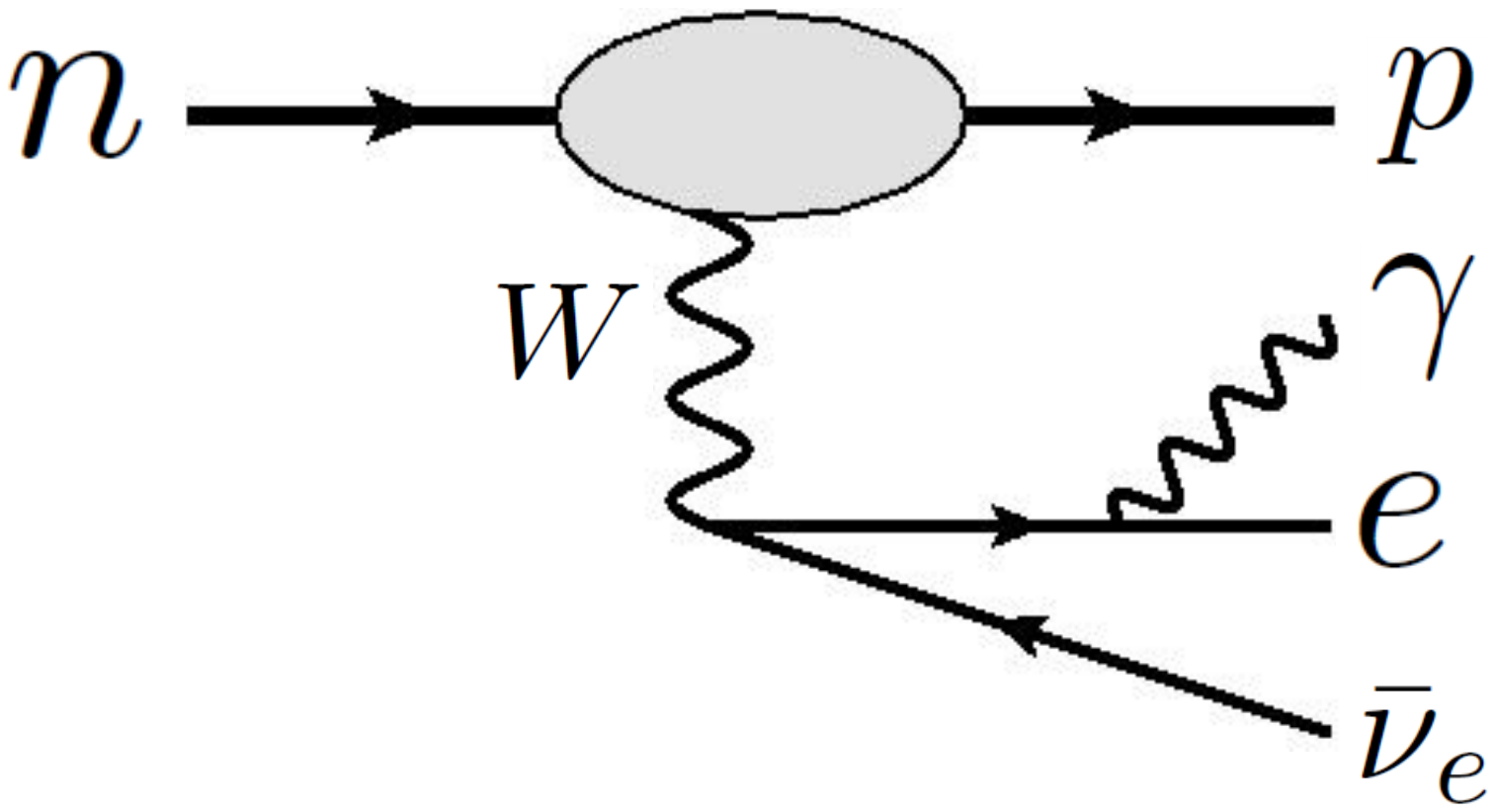}
		\caption{Feynman diagrams for the bremsstrahlung process.}
		\label{fig:brem}
	\end{center}
\end{figure}

Finally, in order to ensure the IR-finiteness of the total decay rate~\cite{Bloch:1937pw,Yennie:1961ad,Kinoshita:1958ru,Kinoshita:1962ur,Lee:1964is}, it is necessary to include the bremsstrahlung process which emits an extra photon, as depicted in Fig.\ref{fig:brem}. The amplitude reads:
\begin{eqnarray}
\mathfrak{M}_\mathrm{brems}&=&-\frac{G_FV_{ud}e}{\sqrt{2}}\frac{1}{2p_e\cdot k}\left\{-2p_e\cdot \varepsilon^*g^{\mu\nu}-k^\mu\varepsilon^{*\nu}+k^\nu\varepsilon^{*\mu}+i\epsilon^{\mu\nu\alpha\beta}k_\alpha\varepsilon_\beta^*\right\}F_\mu L_\nu\nonumber\\
&&+\frac{iG_FV_{ud}e}{\sqrt{2}}\varepsilon^{*\mu}L^\nu T_{\mu\nu}~,\label{eq:Mbrems}
\end{eqnarray}
where $\varepsilon^{*\mu}$ is the polarization vector of the outgoing photon.

\subsection{The three-point function}

It is possible to infer some general properties of the three-point function $\delta F_\text{3pt}^\mu$ just from its definition. We observe that, the first term in Eq.\eqref{eq:3pt} contains an explicit factor of the momentum exchange $q$, while the second term depends on the total derivative of the charged current. This means the first term vanishes in the forward limit ($q\rightarrow 0$) and the second term vanishes in the current conservation limit ($\partial\cdot J_W\rightarrow 0$). 
Due to the small neutron-proton mass splitting $m_n-m_p\approx 1.3$~MeV, recoil effects in the neutron decay cannot be larger than $(m_n-m_p)/m_\pi\sim 10^{-2}$. So, a recoil effect on top of a RC which is already suppressed by $\alpha/\pi\sim 10^{-3}$ gives a $10^{-5}$ corrections which is negligible given our precision goal. Therefore, it is safe to drop the first term in $\delta F_\text{3pt}^\mu$. 

The second term is more tricky, and one should discuss the conservation of the vector and axial component of the charged current separately. The conservation of the vector charged current is only broken by the quark mass difference, i.e. $(\partial\cdot J_W)_V\sim m_u-m_d$; so we again expect a $10^{-5}$ effect which can be safely dropped. This would be the end of the story if the decayed objects are spinless (e.g. pions or spinless nuclei) because only the vector current contributes to $\delta F_\text{3pt}^\mu$, which means we can drop the entire three-point function. But since nucleons have spin, the axial charge current also enters. PCAC tells us that $(\partial\cdot J_W)_A\sim m_\pi^2 P(x)$ where $P(x)$ is the pseudoscalar density~\cite{Gell-Mann:1960mvl}, and one might na\"{\i}vely think the axial current contribution is also suppressed since it is conserved in the chiral limit (i.e. $m_\pi\rightarrow 0$). This is, however, incorrect because the actual separation of scale reads $q\ll m_\pi\ll m_N$, so it is inappropriate to take the chiral limit before the forward limit. Instead, one should  drop terms that scale as $q/m_\pi$ (which is why the pseudoscalar FF is not included in Eq.\eqref{eq:CCME}), and by doing so the axial current becomes explicitly non-conserved. Therefore, there is a non-suppressed contribution to the three-point function from the axial current, and in fact we will find out later that the percent-level correction to $\mathring{g}_A$ discovered in Ref.\cite{Cirigliano:2022hob} is contained entirely in $\delta F_\text{3pt}^\mu$.

\subsection{\label{sec:outin}``Outer'' and ``inner'' corrections}

We may further investigate the properties of the virtual and real RC. The loop integrals in Eqs.\eqref{eq:resintegral}, \eqref{eq:gammaWb} and the bremsstrahlung amplitudes depend only on the tensors $T^{\mu\nu}$ and $\Gamma^\mu$, and it is useful to isolate the ``point-charge'' (pc) contribution which depends only on the electromagnetic and weak charges:
\begin{equation}
T^{\mu\nu}=T^{\mu\nu}_\text{pc}+T^{\mu\nu}_\text{non-pc}~,~\Gamma^{\mu}=\Gamma^{\mu}_\text{pc}+\Gamma^{\mu}_\text{non-pc}~,
\end{equation}
where
\begin{eqnarray}
T_\text{pc}^{\mu\nu}&=&\bar{u}(p_p)\gamma^\mu\frac{i(\slashed{p}_p+\slashed{k}+m_N)}{(p_p+k)^2-m_N^2}\gamma^\nu(\mathring{g}_V+\mathring{g}_A\gamma_5)u(p_n)\nonumber\\
\Gamma^\mu_\text{pc}&=&-2m_N \mathring{g}_A\bar{u}(p_p)\gamma^\mu\frac{\slashed{p}_p+\slashed{k}+m_N}{(p_p+k)^2-m_N^2}\gamma_5u(p_n)~.\label{eq:pc}
\end{eqnarray}
One may check that the pc-contribution satisfies the Ward identities \eqref{eq:Ward} in the isospin limit ($m_p=m_n=m_N$). Physically, the pc-term contributes to the loop integral only at the IR scale: $k\sim q\sim p_e$; the corresponding loop integrations are finite and distorts the beta decay spectrum. Meanwhile the non-pc term probes physics at $k\sim m_\pi$ and above; their only effect is to renormalize the bare coupling constants $\mathring{g}_{V,A}$.  

While the detailed calculation of the three-point function will be discussed in later sections, here we may quote the main conclusion, namely: it splits into an IR-divergent (labeled as ``outer'') and IR-finite (labeled as ``inner'') piece:
\begin{eqnarray}
\delta F_\text{3pt}^\mu&=&(\delta F_\text{3pt}^\mu)^\text{outer}+(\delta F_\text{3pt}^\mu)^\text{inner}\nonumber\\
&=&-\frac{\alpha}{4\pi}\left(\ln\frac{m_\gamma^2}{m_N^2}+2\right)F_A^\mu+\frac{1}{2}\Delta_{R,\text{3pt}}^AF_A^\mu~,\label{eq:3ptsplit}
\end{eqnarray}
the former corrects the beta decay spectrum while the latter introduces a constant $\Delta_{R,\text{3pt}}^A$ that corrects $\mathring{g}_A$. Collecting everything above, we can write the QED-corrected $n\rightarrow pe\bar{\nu}_e$ amplitude as:
\begin{equation}
\mathfrak{M}=\mathfrak{M}_0+\delta\mathfrak{M}=\mathfrak{M}_0+\delta\mathfrak{M}_\text{inner}+\delta\mathfrak{M}_\text{outer}~.\label{eq:Mren}
\end{equation}
The virtual ``outer'' correction arrives by combining the second square bracket in Eq.\eqref{eq:Sirlinrep}, the pc-contribution to Eqs.\eqref{eq:resintegral}, \eqref{eq:gammaWb}, and $(\delta F_\text{3pt}^\mu)^\text{outer}$. After neglecting recoil corrections, the amplitude reads: 
\begin{eqnarray}
\delta\mathfrak{M}_\text{outer}&=&-\frac{G_FV_{ud}}{\sqrt{2}}L_\mu F_\nu\frac{\alpha}{4\pi}\left\{g^{\mu\nu}\left[\frac{1}{2}\ln\frac{m_N^2}{m_e^2}+\ln\frac{m_e^2m_N^2}{m_\gamma^4}-\frac{11}{4}-4p_e\cdot p_pC_0\right]\right.\nonumber\\
&&\left.+\frac{2(\tanh^{-1}\beta-i\pi)(p_p^\mu p_e^\nu-p_e^\mu p_p^\nu-i\epsilon^{\mu\nu\alpha\beta}p_{p\alpha}p_{e\beta})}{p_p\cdot p_e\beta}\right\}~,
\end{eqnarray}
where $\beta=|\vec{p}_e|/E_e$ is the electron speed, and $C_0$ is the standard Passarino-Veltman three-point loop function~\cite{Passarino:1978jh} (see also
Eq.(C.7) in Ref.\cite{Seng:2021wcf}). Meanwhile, the ``inner'' RC simply shifts the vector and axial coupling constants; it combines with the tree-level amplitude to give:
\begin{equation}
\mathfrak{M}_0+\delta \mathfrak{M}_\text{inner}=-\frac{G_FV_{ud}}{\sqrt{2}}L_\lambda\bar{u}(p_p)\gamma^\lambda\left[g_V+g_A\gamma_5\right]u(p_n)~,
\end{equation} 
which defines the ``physical'' $g_V$, $g_A$, the QED-corrected couplings following the notations by Wilkinson and Macefield~\cite{Wilkinson:1970cdv}:
\begin{eqnarray}
	g_V^2&=&\mathring{g}_V^2\left\{1+\Delta_R^V\right\}=\mathring{g}_V^2\left\{1+\Delta_R^U+2\Box_{\gamma W}^V\right\}\nonumber\\
	g_A^2&=&\mathring{g}_A^2\left\{1+\Delta_R^A\right\}=\mathring{g}_A^2\left\{1+\Delta_R^U+2\Box_{\gamma W}^A+2\Box^A_\text{int}+\Delta_{R,\text{3pt}}^{A}\right\}~,
\end{eqnarray}
where $\Delta_\text{R}^U$ is the ``universal'' inner RC in Eq.\eqref{eq:universal}, $\Box_{\gamma W}^{V,A}$ come from the non-pc contribution to $\delta \mathfrak{M}_{\gamma W}^b$ and can be studied to high precision using dispersion relation~\cite{Seng:2018yzq,Seng:2018qru,Seng:2020wjq,Shiells:2020fqp} and lattice QCD~\cite{Ma:2023kfr}, $\Box^A_\text{int}$ comes from the contribution of $\Gamma^\mu_\text{non-pc}$ to the ``residual integral'', and $\Delta_{R,\text{3pt}}^A$ comes from $\delta F_\text{3pt}^\mu$ which contributes only to the axial FF. Notice that $\Delta_R^V$ is the famous ``nucleus-independent RC'' that affects the determination of $V_{ud}$ from neutron and nuclear decay processes~\cite{Hardy:2020qwl}. From the equation above we also obtain the following QED-corrected axial-to-vector ratio:
\begin{equation}
\lambda^2=\frac{g_A^2}{g_V^2}=\frac{\mathring{g}_A^2}{\mathring{g}_V^2}\left\{1+2(\Box_{\gamma W}^A-\Box_{\gamma W}^V)+2\Box^A_\text{int}+\Delta_{R,\text{3pt}}^A\right\}~.\label{eq:ratio}
\end{equation}
Previous works such as \cite{Gorchtein:2021fce} missed the terms $\Box^A_\text{int}$ and  $\Delta_{R,\text{3pt}}^A$. 

To evaluate the bremsstrahlung amplitude, we notice that it depends only on physics at the scale $k\sim q\sim p_e$ since the photon must be on-shell, thus it is appropriate to replace $T^{\mu\nu}\rightarrow T_\text{pc}^{\mu\nu}$. Adding everything above, we are now able to write down the differential rate of the inclusive decay process $n\rightarrow pe\bar{\nu}_e(\gamma)$ for a polarized neutron as: 
\begin{eqnarray}
\frac{d\Gamma}{dE_ed\Omega_ed\Omega_\nu}&=&\frac{(G_FV_{ud})^2}{(2\pi)^5}|\vec{p}_e|E_e(E_m-E_e)^2F(E_e)g_V^2(1+3\lambda^2)\left(1+\frac{\alpha}{2\pi}g(E_e)\right)\nonumber\\
&&\times\left\{1+\left(1+\frac{\alpha}{2\pi}\delta^{(2)}(E_e)\right)a\vec{\beta}\cdot\hat{p}_\nu+\hat{e}_s\cdot\left[\left(1+\frac{\alpha}{2\pi}\delta^{(2)}(E_e)\right)A\vec
\beta+B\hat{p}_\nu\right]\right\}~,\nonumber\\\label{eq:differential}
\end{eqnarray}
where $\beta=\vec{p}_e/E_e$, $\hat{e}_s$ is the neutron unit spin vector, $E_m=(m_n^2-m_p^2+m_e^2)/(2m_n)\approx m_n-m_p$ is the electron end-point energy, $F(E_e)\approx 1+\alpha\pi/\beta$ is the Fermi function~\cite{Fermi:1934hr} (see also more recent discussions in ~\cite{Cirigliano:2023fnz,Hill:2023acw,Hill:2023bfh}), and
\begin{equation}
a=\frac{1-\lambda^2}{1+3\lambda^2}~,~A=-\frac{2\lambda(\lambda+1)}{1+3\lambda^2}~,~B=\frac{2\lambda(\lambda-1)}{1+3\lambda^2}~
\end{equation}
are the correlation coefficients that can be experimentally measured. The functions $g(E_e)$ and $\delta^{(2)}(E_e)$ describe the full outer corrections (after excluding the Fermi function), and are given by~\cite{Sirlin:1967zza,Garcia:1981it}:
\begin{eqnarray}
g(E_e)&=&3\ln\frac{m_N}{m_e}-\frac{3}{4}+4\left(\frac{1}{\beta}\tanh^{-1}\beta-1\right)\left(\ln\frac{2(E_m-E_e)}{m_e}+\frac{E_m-E_e}{3E_e}-\frac{3}{2}\right)\nonumber\\
&&-\frac{4}{\beta}\mathrm{Li}_2\left(\frac{2\beta}{1+\beta}\right)+\frac{1}{\beta}\tanh^{-1}\beta\left(2+2\beta^2+\frac{(E_m-E_e)^2}{6E_e^2}-4\tanh^{-1}\beta\right)\nonumber\\
\delta^{(2)}(E_e)&=&2\left(\frac{1-\beta^2}{\beta}\right)\tanh^{-1}\beta+\frac{4(E_m-E_e)(1-\beta^2)}{3\beta^2E_e}\left(\frac{1}{\beta}\tanh^{-1}\beta-1\right)\nonumber\\
&&+\frac{(E_m-E_e)^2}{6\beta^2E_e^2}\left(\frac{1-\beta^2}{\beta}\tanh^{-1}\beta-1\right)~.
\end{eqnarray}
In particular, $g(E_e)$ is the well-known Sirlin's function~\cite{Sirlin:1967zza}. It is worth pointing out that, despite its  mathematical rigorosity,  Eq.\eqref{eq:differential} is actually problematic when applied to the analysis of actual experimental outputs, especially the $\vec{p}_\nu$-dependent correlation functions. This is because most experiments do not detect the outgoing neutrino, so it is the following ``pseudo-neutrino'' momentum:
\begin{equation}
	\vec{p}_\nu'\equiv -\vec{p}_p-\vec{p}_e~,
\end{equation}
instead of $\vec{p}_\nu$, which is an actual experimental observable. The equality $\vec{p}_\nu'=\vec{p}_\nu$ breaks down in the presence of the bremsstrahlung photon, so the na\"{\i}ve application of Eq.\eqref{eq:differential} would lead to an $\mathcal{O}(\alpha/\pi)$ error in the $\vec{p}_e$-dependent correlations. A possible solution is to use $\vec{p}_\nu'$ in replacement of $\vec{p}_\nu$ as a free variable in the differential rate; by doing so one modifies Eq.\eqref{eq:differential} as:
\begin{eqnarray}
	\frac{d\Gamma}{dE_ed\Omega_ed\Omega_\nu'}&=&\frac{(G_F V_{ud})^2}{(2\pi)^5}|\vec{p}_e|E_e(E_m-E_e)^2F(E_e)g_V^2(1+3\lambda^2)\Bigg\{1+a\beta c'+\frac{\delta_\text{tot}^0(E_e,c')}{1+3\lambda^2}\nonumber\\
	&&+\hat{e}_s\cdot\left[\left(A+\frac{\delta_\text{tot}^\beta(E_e,c')}{1+3\lambda^2}\right)\vec{\beta}+\left(B+\frac{\delta_\text{tot}^\nu(E_e,c')}{1+3\lambda^2}\right)\hat{p}_\nu'\right]\Bigg\}~,
\end{eqnarray}
where $c'\equiv \hat{p}_e\cdot\hat{p}_\nu'$, and the analytic expressions of $\delta_\text{tot}^{0,\beta,\nu}$ can be found in Ref.\cite{Seng:2023ynd}. Powerful numerical packages that return equivalent results are also available~\cite{Gluck:2022ogz}. 

\section{\label{sec:residual}Inner correction from the residual integral}

Given that the quantities $\Box_{\gamma W}^{V,A}$ are now well-understood, we need only to determine $\Box^A_\text{int}$ and $\Delta^A_{R,\text{3pt}}$ to fully pin down the inner RCs, and we start with $\Box^A_\text{int}$ which originates from the ``residual integral''. Among the four terms in the curly bracket of Eq.\eqref{eq:resintegral}, the first three terms have an explicit suppression of $p_e$ or $p_n-p_p$ in the numerator, therefore their integrals can only depend on physics at the scale $k\sim p_e$, which means the replacement $T^{\mu\nu}\rightarrow T^{\mu\nu}_\text{pc}$ is sufficient. This is, however, not true for the fourth term: it depends not only on physics at $k\sim p_e$ (which is evaluated by setting $\Gamma^\lambda\rightarrow \Gamma^\lambda_\text{pc}$), but also up to $k\sim m_\pi$ where the pseudoscalar form factor $G_P$ cannot be neglected; it does not probe the scale $k\gg m_\pi$ where chiral symmetry is restored and $\Gamma^\lambda\rightarrow 0$. Therefore, to compute the full contribution of the fourth term in Eq.\eqref{eq:resintegral} we need to include all the nucleon electromagnetic and weak form factors (here we take $p_p=p_n=p$ for simplicity):
\begin{eqnarray}
	\Gamma^\mu&=&-2m_N G_A\frac{m_\pi^2}{m_\pi^2-k^2}\bar{u}_p\left\{\frac{[F_{1p}\gamma^\mu-\frac{i}{2m_N}\sigma^{\mu\nu}k_\nu F_{2p}](\slashed{p}+\slashed{k}+m_N)\gamma_5}{(p+k)^2-m_N^2}\right.\nonumber\\
	&&\left.+\frac{\gamma_5(\slashed{p}-\slashed{k}+m_N)[F_{1n}\gamma^\mu-\frac{i}{2m_N}\sigma^{\mu\nu}k_\nu F_{2n}]}{(p-k)^2-m_N^2}\right\}u_n~,\label{eq:Gammafull}
\end{eqnarray}
and $\Gamma^\mu_{\text{pc}}$ in Eq.\eqref{eq:pc} simply corresponds to taking $k\rightarrow 0$ everywhere except the nucleon propagators in the expression above. 

Combining Eqs.\eqref{eq:pc} and \eqref{eq:Gammafull}, we obtain the analytic expression of $\Gamma^\nu_\text{non-pc}=\Gamma^\mu-\Gamma^\mu_{\text{pc}}$ which is then plugged into Eq.\eqref{eq:resintegral} (and take $p_e\rightarrow 0$) to get its inner RC to $g_A$. The result takes the following integral form:
\begin{eqnarray}
	\Box^A_\text{int}&=&-\frac{\alpha}{2\pi}\int_0^\infty\frac{dQ^2}{Q^2}\left\{\frac{Q^2}{4m_N^2}\left(\frac{2m_N^2}{Q^2}+1-\sqrt{1+\frac{4m_N^2}{Q^2}}\right)\times\right.\nonumber\\
	&&\left[\left(F_{1p}(Q^2)-F_{1n}(Q^2)\right)\hat{G}_A(Q^2)\frac{m_\pi^2}{m_\pi^2+Q^2}-1\right]-\frac{1}{3}\left(F_{2p}(Q^2)-F_{2n}(Q^2)\right)\times\nonumber\\
	&&\left.\hat{G}_A(Q^2)\frac{m_\pi^2}{m_\pi^2+Q^2}\left[\frac{(Q^2)^2}{16m_N^4}-\frac{3Q^2}{8m_N^2}+\left(\frac{Q^2}{2m_N^2}-\frac{(Q^2)^2}{16m_N^4}\right)\sqrt{1+\frac{4m_N^2}{Q^2}}\right]\right\}~,
\end{eqnarray}
where $\hat{G}_A(Q^2)\equiv G_A(Q^2)/\mathring{g}_A$. The integral can be evaluated analytically by taking the $Q^2\rightarrow 0$ limit to all the form factors: $F_{1p}(0)=\hat{G}_A(0)=1$, $F_{1n}(0)=0$, $F_{2p}(0)=\kappa_p=\mu_p-1$, $F_{2n}(0)=\kappa_n=\mu_n$. The result is:
\begin{eqnarray}
	\Box^A_\text{int}&=&-\frac{\alpha}{4\pi}\left\{\Lambda-\frac{2m_N^2-m_\pi^2}{2m_N^2}\ln\frac{m_N^2}{m_\pi^2}+1\right.\nonumber\\
	&&\left.-\frac{1-\mu_p+\mu_n}{24m_N^4}\left[(16m_N^4+2m_N^2m_\pi^2)\Lambda+(6m_N^2m_\pi^2+m_\pi^4)\ln\frac{m_N^2}{m_\pi^2}+2m_N^2m_\pi^2\right]\right\}\nonumber\\
	&\approx&2.47\times 10^{-3}~,
\end{eqnarray}
where $\Lambda\equiv (m_\pi/m_N^2)\sqrt{m_\pi^2-4m_N^2}\ln[(\sqrt{m_\pi^2-4m_N^2}+m_\pi)/(2m_N)]$. We also checked that, retaining the full nucleon form factor-dependence~\cite{Drechsel:2007if,Lorenz:2012tm,Lorenz:2014yda,Ye:2017gyb,Lin:2021umk,Lin:2021umz,Bhattacharya:2011ah} only shifts its value slightly to $2.44\times 10^{-3}$ with an uncertainty of the order $10^{-6}$ from the form factors; this proves that the $\Gamma^\mu$ contribution is indeed insensitive to physics at $k\gg m_\pi$. For the later comparison with EFT, it is also useful to obtain its expansion in powers of $m_\pi/m_N$:
\begin{equation}
	\Box^A_\text{int}=\frac{\alpha}{2\pi}\left\{\frac{1}{2}\left(\ln\frac{m_N^2}{m_\pi^2}-1\right)+\frac{\pi m_\pi}{6m_N}(1+2\mu_p-2\mu_n)+\mathcal{O}\left(\frac{m_\pi^2}{m_N^2}\right)\right\}~.\label{eq:Boxintexpand}
\end{equation}

\section{\label{sec:EFT}Introducing the effective field theory framework}

We now deal with  $\Delta_{R,\text{3pt}}^A$ that involves a three-current product and is more difficult to be treated with the method above. We resort to the hybrid analysis introduced in Ref.\cite{Seng:2019lxf}, namely to compute not the full amplitude, but only the FF correction $\delta F^\mu$ in an EFT framework, and perform manually the splitting of two- and three-point function. This provides an EFT prediction of $\Delta_{R,\text{3pt}}^A$ which can be used alongside with other, more accurately-predicted inner RC. For that purpose, we need to introduce the EFT Lagrangian from which the Feynman rules of the charged weak current operator and other QCD+QED interactions are derived. In this work we stick to the HBChPT framework adopted in Ref.\cite{Cirigliano:2022hob}, with some minor changes of notations.  

We start by introducing external source terms to the two-flavor massless QCD Lagrangian $\mathcal{L}_\text{QCD}^0$:
\begin{equation}
\mathcal{L}=\mathcal{L}_\text{QCD}^0+\bar{q}_R\gamma_\mu(r^\mu+v_s^\mu/3)q_R+q_L\gamma_\mu(l^\mu+v_s^\mu/3)q_L-\bar{q}_R(s+ip)q_L-\bar{q}_L(s-ip)q_R~,
\end{equation}
where $r^\mu$, $l^\mu$, $v_s^\mu$, $s$ and $p$ are external Hermitian source fields. In particular, 
$r^\mu$, $l^\mu$ are isotriplet (i.e. traceless in the flavor space) while $v_s^\mu$ is isosinglet. To compute $\delta F^\mu$ we need to include dynamical photons but not leptons.
This amounts to splitting the external fields as:
\begin{eqnarray}
r^\mu&=&-eA_\mu\left\{Q_\text{EM}^R-\frac{1}{2}\langle Q_\text{EM}^R\rangle\right\}+\bar{r}^\mu\nonumber\\
l^\mu&=&-eA_\mu\left\{Q_\text{EM}^L-\frac{1}{2}\langle Q_\text{EM}^L\rangle\right\}+\bar{l}^\mu\nonumber\\
v_s^\mu&=&-\frac{3e}{4}A_\mu\langle Q_\text{EM}^R+Q_\text{EM}^L\rangle +\bar{v}_s^\mu\nonumber\\
s+ip&=&M_q+\bar{s}+i\bar{p}~,
\end{eqnarray}
where $Q_\text{EM}^R=Q_\text{EM}^L=\text{diag}(2/3,-1/3)$ is the quark charge matrix, $M_q=\text{diag}(m_u,m_d)$ is the quark mass matrix, $A_\mu$ is the photon field, and $\langle...\rangle$ denotes the trace over flavor space. Throughout this work we will assume strong isospin symmetry, i.e. $m_u=m_d\equiv m_q$. The physical Lagrangian is recovered when the ``classical fields'' $\bar{r}^\mu$, $\bar{l}^\mu$, $\bar{v}_s^\mu$, $\bar
s$ and $\bar{p}$ are set to zero. Particular interest is on the field $\bar{l}^\mu$ attached to the left-handed current: by spelling out its isotriplet components,
\begin{equation}
\bar{l}_\mu=2\tau_+\bar{l}^+_\mu+2\tau_-\bar{l}^-_\mu+\tau_3\bar{l}^0_\mu~,
\end{equation}
the charged current operator $J_W^\mu$ can be obtained as:
\begin{equation}
J_W^\mu=\left(\frac{\partial}{\partial\bar{l}^+_\mu}\mathcal{L}\right)_{\bar{l}_\mu=0}~.
\end{equation}
The equation above allows us to derive the Feynman rules of the $J_W^\mu$ vertices in an EFT. 

Now we proceed to introduce the chiral Lagrangian; standard notations in HBChPT can be found in Ref.\cite{Scherer:2012xha} and will not be repeated here. We assign the chiral power counting $e\sim p$ which allows us to retain finite terms in the full chiral Lagrangian given our precision goal. In the pion sector we only need the Lagrangian to the chiral order $n=2$:
\begin{equation}
\mathcal{L}_\pi=\mathcal{L}_\pi^{(2)}+...=\mathcal{L}_\pi^{p^2}+\mathcal{L}_\pi^{e^2}+...~,
\end{equation}
where
\begin{equation}
\mathcal{L}_\pi^{p^2}=\frac{F_0^2}{4}\langle u_\mu u^\mu+\chi_+\rangle~,~\mathcal{L}_\pi^{e^2}=e^2Z_\pi F_0^4\langle Q_\text{EM}^R U Q_\text{EM}^L U^\dagger\rangle~,
\end{equation}
with
\begin{eqnarray}
u_\mu&=&i\left[u^\dagger(\partial_\mu-ir_\mu)u-u(\partial_\mu-il_\mu)u^\dagger\right]\nonumber\\
\chi_+&=&u^\dagger\chi u^\dagger+u\chi^\dagger u\nonumber\\
\chi&=&B_0(s+ip)~.
\end{eqnarray}
In particular, the $\mathcal{O}(e^2)$ term describe short-distance QED effects where the photons are integrated out~\cite{Ecker:1988te}. The LEC $Z_\pi\approx 0.81$ is obtained from $\pi^+-\pi^0$ mass splitting.  

In the nucleon sector, we need the chiral Lagrangian up to $n=3$:
\begin{equation}
\mathcal{L}_{\pi N}=\mathcal{L}_{\pi N}^{(1)}+\mathcal{L}_{\pi N}^{(2)}+\mathcal{L}_{\pi N}^{(3)}+...
\end{equation}
The $n=1$ Lagrangian is well-known:
\begin{equation}
\mathcal{L}_{\pi N}^{(1)}=\bar{N}_v iv\cdot \mathcal{D}N_v-\mathring{g}_A \bar{N}_vS\cdot u N_v~,
\end{equation}
where the chiral covariant derivative on the nucleon field reads:
\begin{equation}
\mathcal{D}_\mu N_v=\left(\partial_\mu+\frac{1}{2}\left[u^\dagger(\partial_\mu-ir_\mu)u+u(\partial_\mu-il_\mu)u^\dagger\right]-iv_\mu^{s}\right)N_v~.
\end{equation}
At $n=2$ there are two terms:
\begin{equation}
\mathcal{L}_{\pi N}^{(2)}=\mathcal{L}_{\pi N}^{p^2}+\mathcal{L}_{\pi N}^{e^2}~.
\end{equation}
The $\mathcal{O}(e^2)$ term gives short-distance QED corrections to nucleon masses, which effect is canceled out in the full amplitude. So we display only the $\mathcal{O}(p^2)$ term:
\begin{eqnarray}
	\mathcal{L}_{\pi N}^{p^2}&=&\bar{N}_v\left[\frac{1}{2m_N}\left((v\cdot\mathcal{D})^2-\mathcal{D}^2\right)+\frac{i\mathring{g}_A}{2m_N}\{S\cdot\mathcal{D},u\cdot v\}+c_1\langle\chi_+\rangle\right.\nonumber\\
	&&+\left(c_2+\frac{\mathring{g}_A}{8m_N}\right)(u\cdot v)^2+c_3u\cdot u+\left(c_4+\frac{1}{4m_N}\right)[S^\mu,S^\nu]u_\mu u_\nu+c_5\tilde{\chi}_+\nonumber\\
	&&\left.-\frac{i}{4m_N}[S^\mu,S^\nu]\left((1+\mathring{\kappa}_V)f^+_{\mu\nu}+2(1+\mathring{\kappa}_S)v_{\mu\nu}^{s}\right)\right]N_v~,
\end{eqnarray}
where 
\begin{eqnarray}
\tilde{\chi}_+&=&\chi_+-\frac{1}{2}\langle\chi_+\rangle\nonumber\\
f_{\mu\nu}^+&=&u\left(\partial_\mu l_\nu-\partial_\nu l_\mu-i[l_\mu,l_\nu]\right)u^\dagger+u^\dagger\left(\partial_\mu r_\nu-\partial_\nu r_\mu-i[r_\mu,r_\nu]\right)u\nonumber\\
v_{\mu\nu}^{s}&=&\partial_\mu v_\nu^{s}-\partial_\nu v_\mu^{s}~,
\end{eqnarray} $\mathring{\kappa}_S=\mathring{\kappa}_p+\mathring{\kappa}_n$, $\mathring{\kappa}_V=\mathring{\kappa}_p-\mathring{\kappa}_n$ are the isoscalar and isovector nucleon magnetic moments in the chiral limit (the physical magnetic moments are $\kappa_p\approx 1.793$, $\kappa_n\approx -1.913$), and the LECs $c_i$ can be extracted from pion-nucleon scattering~\cite{Hoferichter:2015hva,Siemens:2016jwj}. Notice that the magnetic moment terms here appear to be different from that in Ref.\cite{Cirigliano:2022hob} because we use the quark charge matrix, instead of the nucleon charge matrix, as the spurion field. Finally, at $n=3$ there are also two terms:
\begin{equation}
\mathcal{L}_{\pi N}^{(3)}=\mathcal{L}_{\pi N}^{p^3}+\mathcal{L}_{\pi N}^{e^2p}~,
\end{equation}
but we only need the part in $\mathcal{L}_{\pi N}^{e^2p}$ that contributes to $F_A^\mu$ at tree level:
\begin{equation}
\mathcal{L}_{\pi N}^{e^2p}=-2e^2\left(g_1+g_2+\frac{g_{11}}{2}\right)\bar{N}_vS^\mu\left(\tau_+\bar{l}_\mu^++\tau_-\bar{l}_\mu^-\right)N_v+...\label{eq:counter}
\end{equation}
The LECs $\{g_i\}$ are defined in Ref.\cite{Gasser:2002am}, except that we rescale them as $F_0^2(g_i)_\text{there}=(g_i)_\text{here}$ for convenience. The above are the full chiral Lagrangian which our following EFT analysis is based on.

\section{\label{sec:HBChPT}Radiative corrections to the axial form factor}

\begin{figure}[t]
	\begin{center}
		\includegraphics[width=0.22\columnwidth]{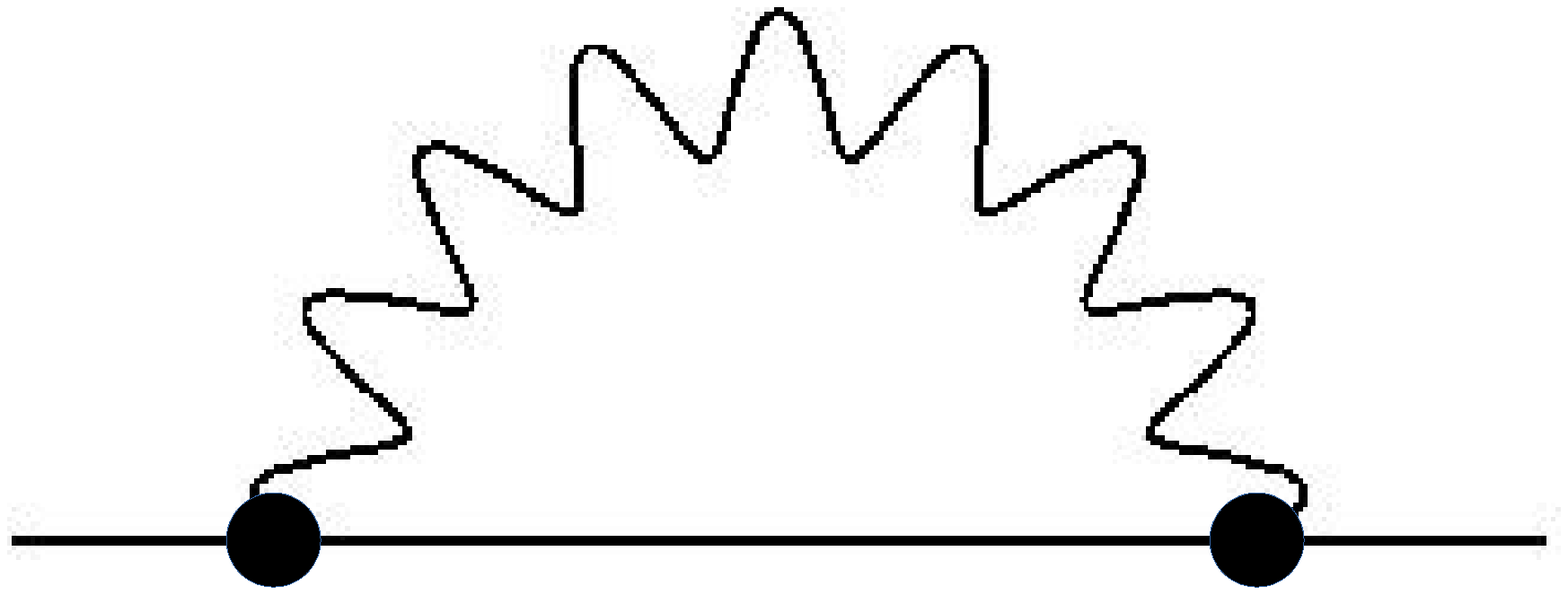}
		\includegraphics[width=0.22\columnwidth]{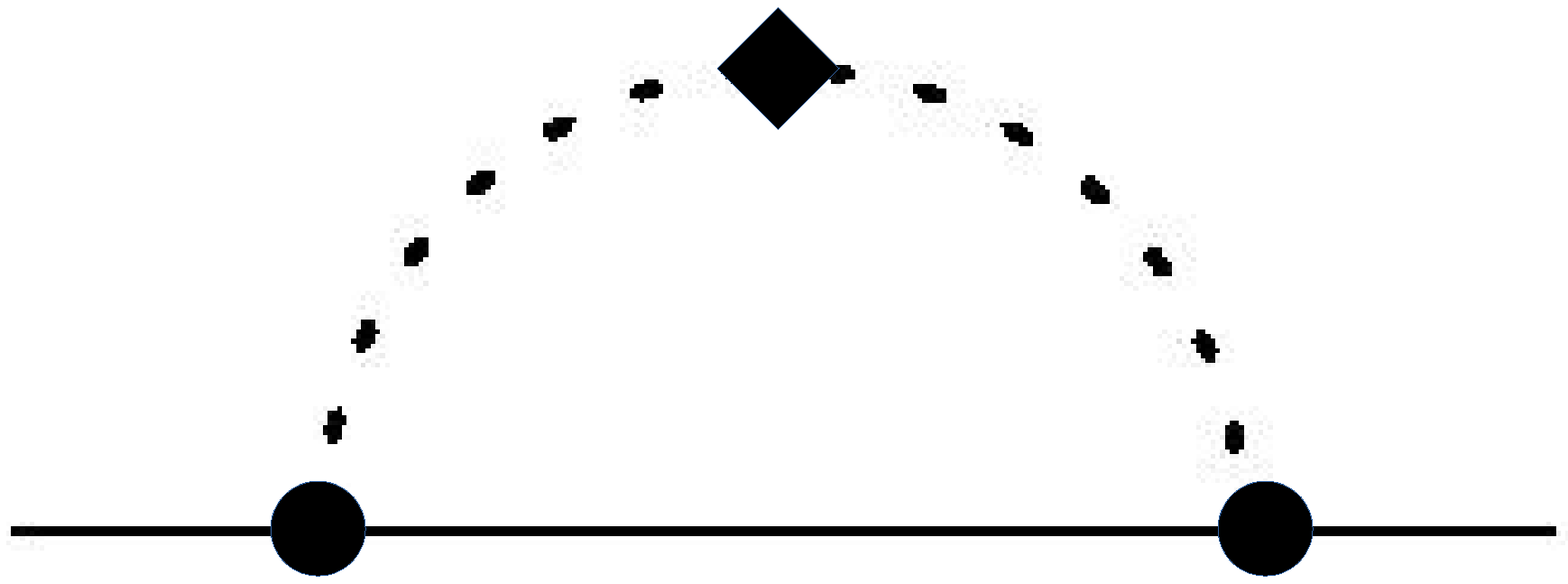}
		\includegraphics[width=0.22\columnwidth]{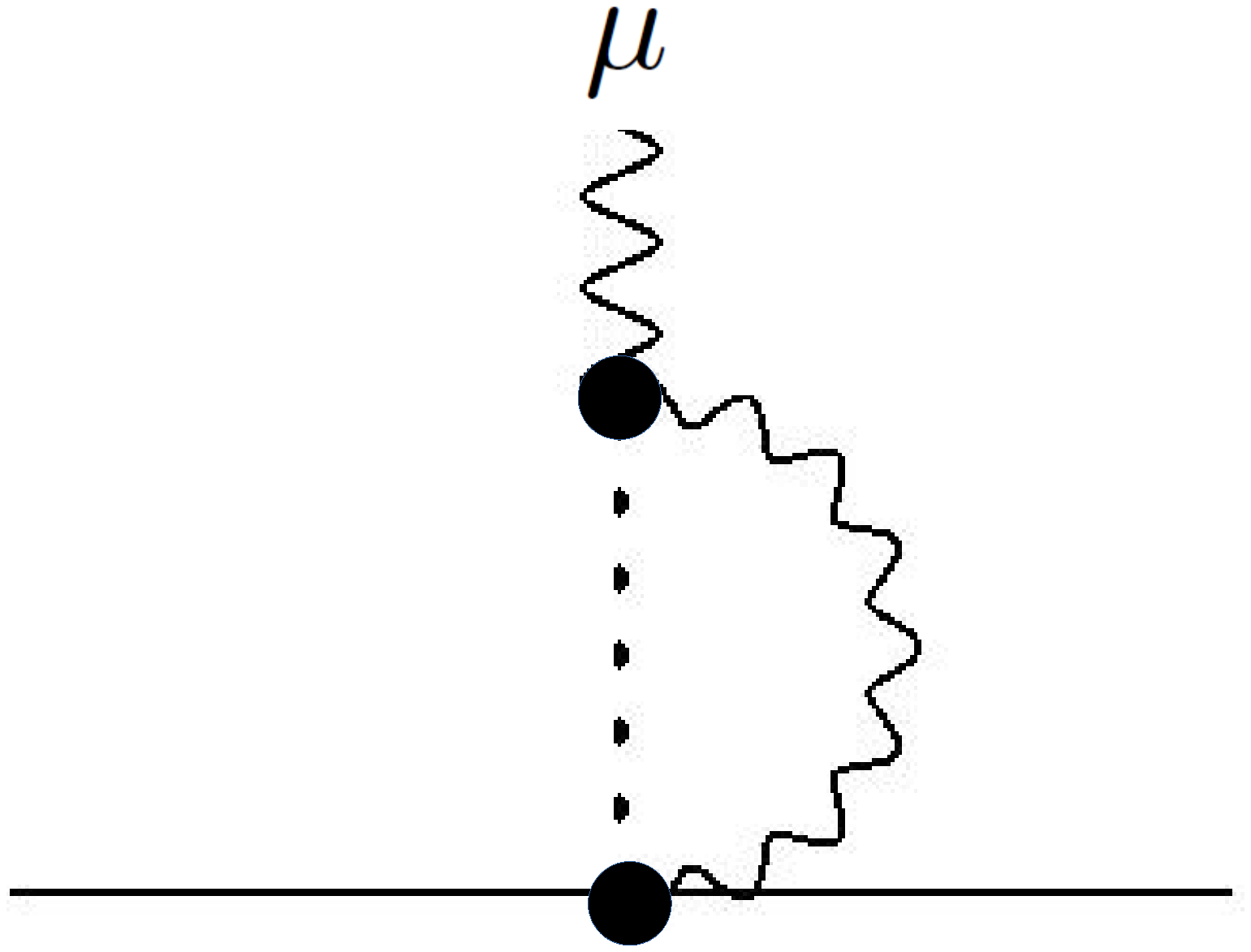}
		\includegraphics[width=0.22\columnwidth]{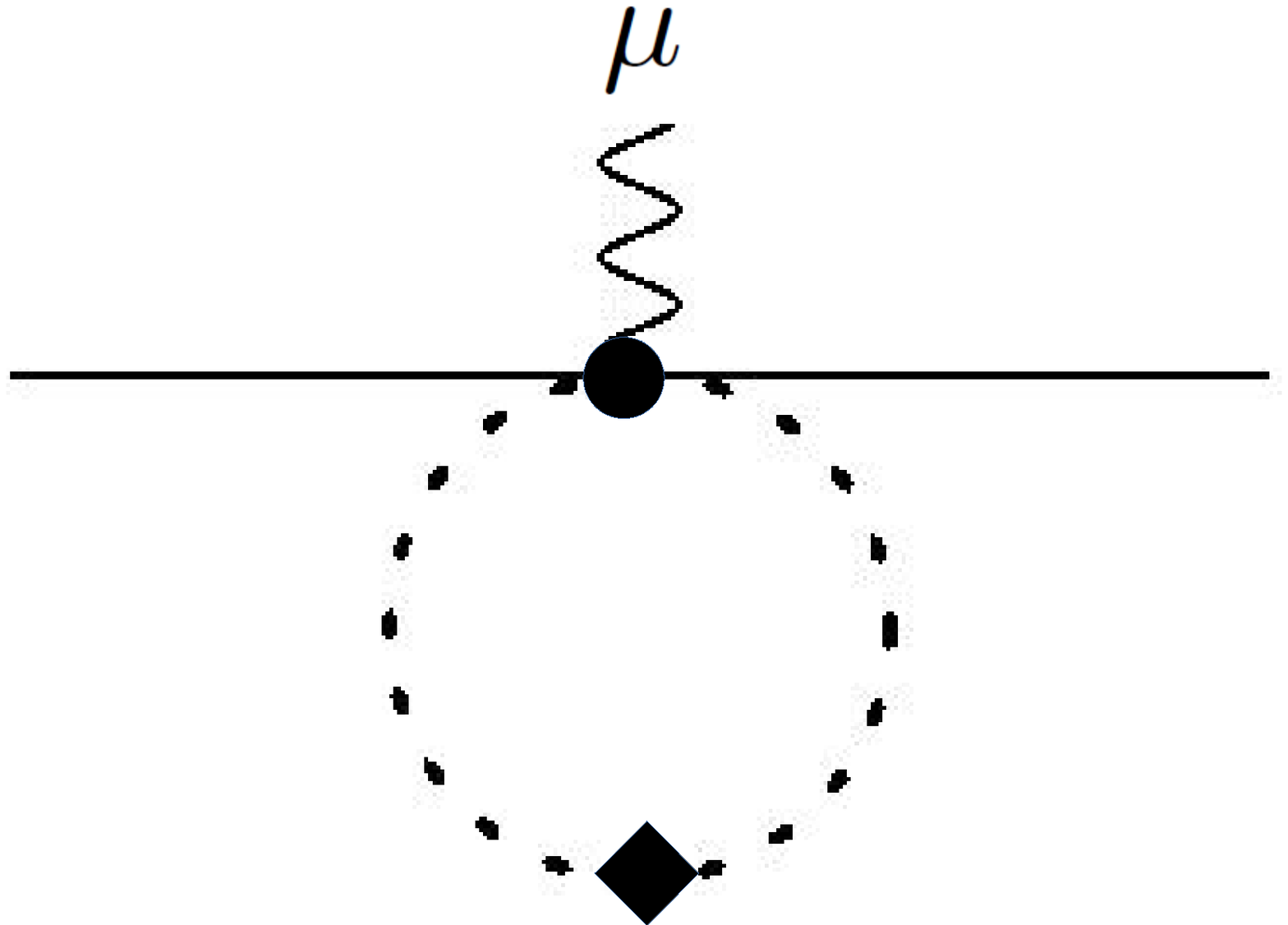}
		\caption{LO loop diagrams, both self-energy and one-particle-irreducible (1PI), for $\delta F_A^\mu$. Solid line: nucleon. Dashed line: pion. Internal wavy line: photon. External wavy line: connection to $\bar{l}^+_\mu$. Dot: vertices from  $\mathcal{L}_\pi^{p^2}$ and $\mathcal{L}_{\pi N}^p$. Diamond: vertex from $\mathcal{L}_\pi^{e^2}$.   }
		\label{fig:LO}
	\end{center}
\end{figure}

Now we are in the position to apply the EFT framework. As discussed before, the vector FF correction is fully under control, so we may concentrate on the RC to the axial FF, $\delta F_A^\mu$. There are two steps in our analysis: We first compute the full $\delta F_A^\mu$ in EFT, and then separate it into two- and three-point function; with this we can infer the value of $\Delta_{R,\text{3pt}}^A$. 

\subsection{Full $\delta F_A^\mu$}

The lowest-order $F_A^\mu$ is derived from $\mathcal{L}_{\pi N}^{p}$ and thus scales as $\mathcal{O}(p)$; we denote the terms in the FF correction $\delta F_A^\mu$ that scale as $\mathcal{O}(e^2p)$ and $\mathcal{O}(e^2p^2)$ as the ``leading-order'' (LO) and ``next-to-leading-order'' (NLO) corrections, respectively. One-loop diagrams at LO are given in Fig.\ref{fig:LO}; the loop integrals contain ultraviolet (UV)-divergences which are canceled by the $\mathcal{O}(e^2p)$ LECs in Eq.\eqref{eq:counter}. The result reads:
\begin{eqnarray}
	(\delta F_A^\mu)_\text{LO}&=&\frac{\alpha}{2\pi}\left[\frac{Z_\pi}{2}(1+3\mathring{g}_A^2)\left(\ln\frac{\mu^2}{m_\pi^2}-1\right)-\mathring{g}_A^2Z_\pi-\frac{1}{2}\left(\ln\frac{m_\gamma^2}{m_\pi^2}+1\right)\right.\nonumber\\
	&&\left.-\frac{8\pi^2}{\mathring{g}_A}\left(g_1^r+g_2^r+\frac{g_{11}^r}{2}\right)\right]F_A^\mu~,
\end{eqnarray}
where $\mu$ is the renormalization scale in the dimensional-regularization prescription. The renormalized LECs $g_i^r$ are related to the bare LECs $g_i$ as:
\begin{equation}
g_i=\eta_i\frac{\mu^{d-4}}{(4\pi)^2}\left(\frac{1}{d-4}-\frac{1}{2}\left(-\gamma+\ln 4\pi+1\right)\right)+g_i^r(\mu)~.
\end{equation}
The values of the coefficients $\eta_i$ can be found in Ref.\cite{Gasser:2002am}.
As we advocate in Sec.\ref{sec:outin}, one can separate the expression above into ``outer'' and ``inner'' piece:
\begin{equation}
(\delta F_A^\mu)_\text{LO}=(\delta F_A^\mu)^\text{outer}+(\delta F_A^\mu)_\text{LO}^\text{inner}~,
\end{equation}
where
\begin{eqnarray}
(\delta F_A^\mu)^\text{outer}&=&-\frac{\alpha}{4\pi}\left(\ln\frac{m_\gamma^2}{m_N^2}+2\right)F_A^\mu\nonumber\\
(\delta F_A^\mu)_\text{LO}^\text{inner}&=&\frac{\alpha}{2\pi}\left[\frac{Z_\pi}{2}(1+3\mathring{g}_A^2)\left(\ln\frac{\mu^2}{m_\pi^2}-1\right)-\mathring{g}_A^2Z_\pi-\frac{1}{2}\left(\ln\frac{m_N^2}{m_\pi^2}-1\right)\right.\nonumber\\
&&\left.-\frac{8\pi^2}{\mathring{g}_A}\left(g_1^r+g_2^r+\frac{g_{11}^r}{2}\right)\right]F_A^\mu~.
\end{eqnarray} 

\begin{figure}[t]
	\begin{center}
		\includegraphics[width=0.22\columnwidth]{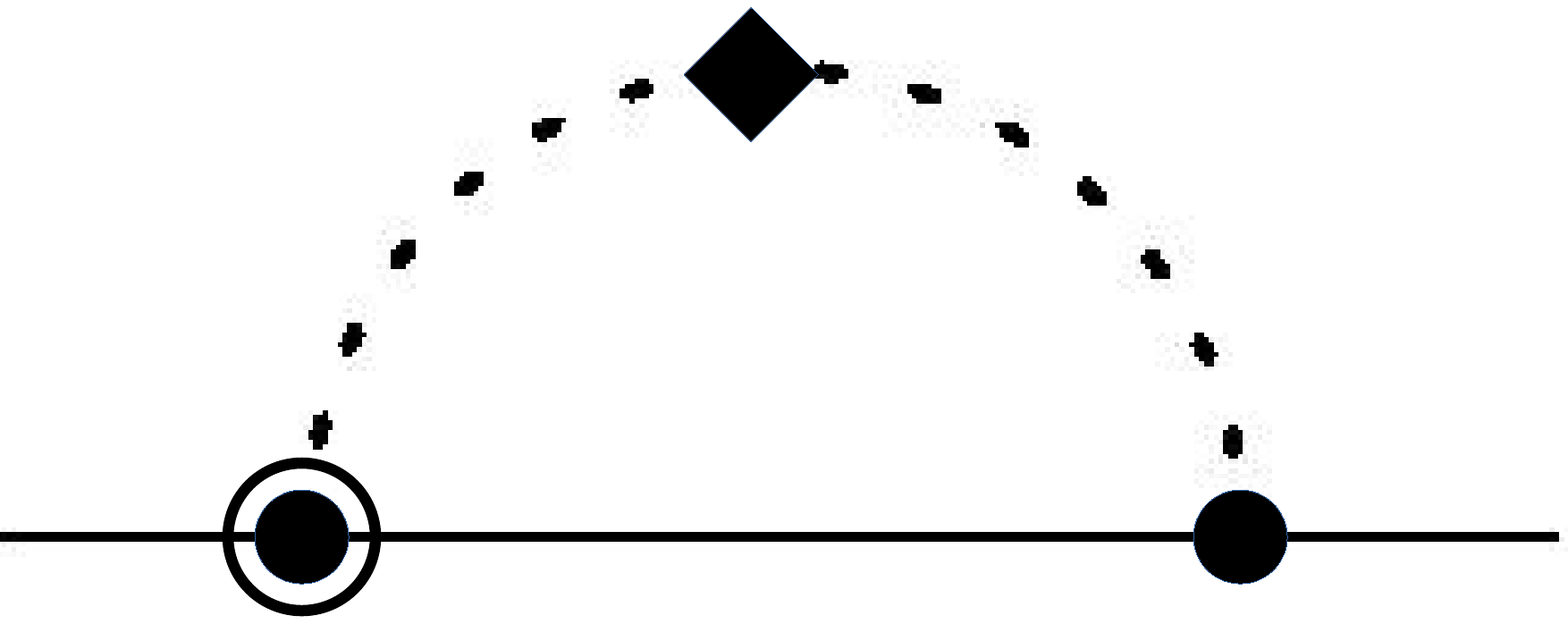}
		\includegraphics[width=0.22\columnwidth]{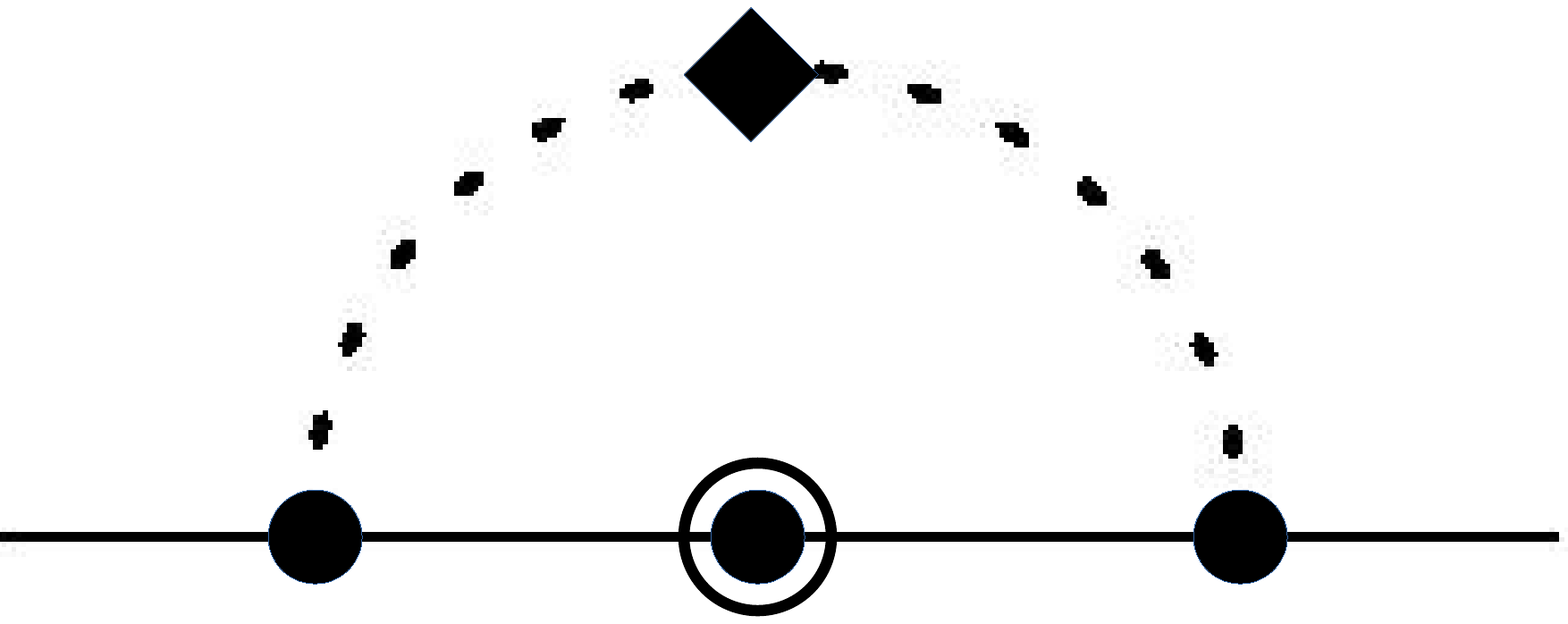}
		\includegraphics[width=0.22\columnwidth]{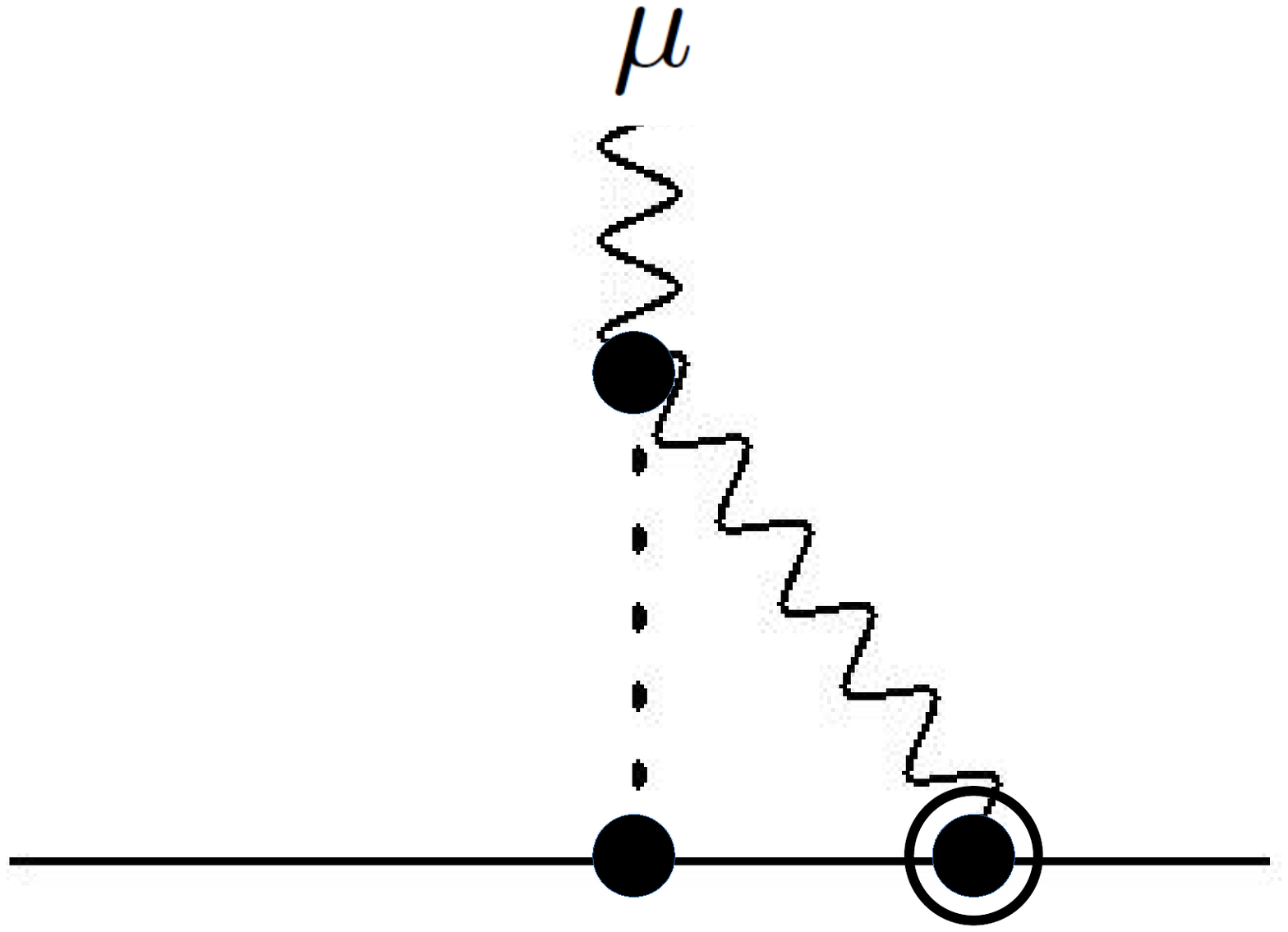}
		\includegraphics[width=0.22\columnwidth]{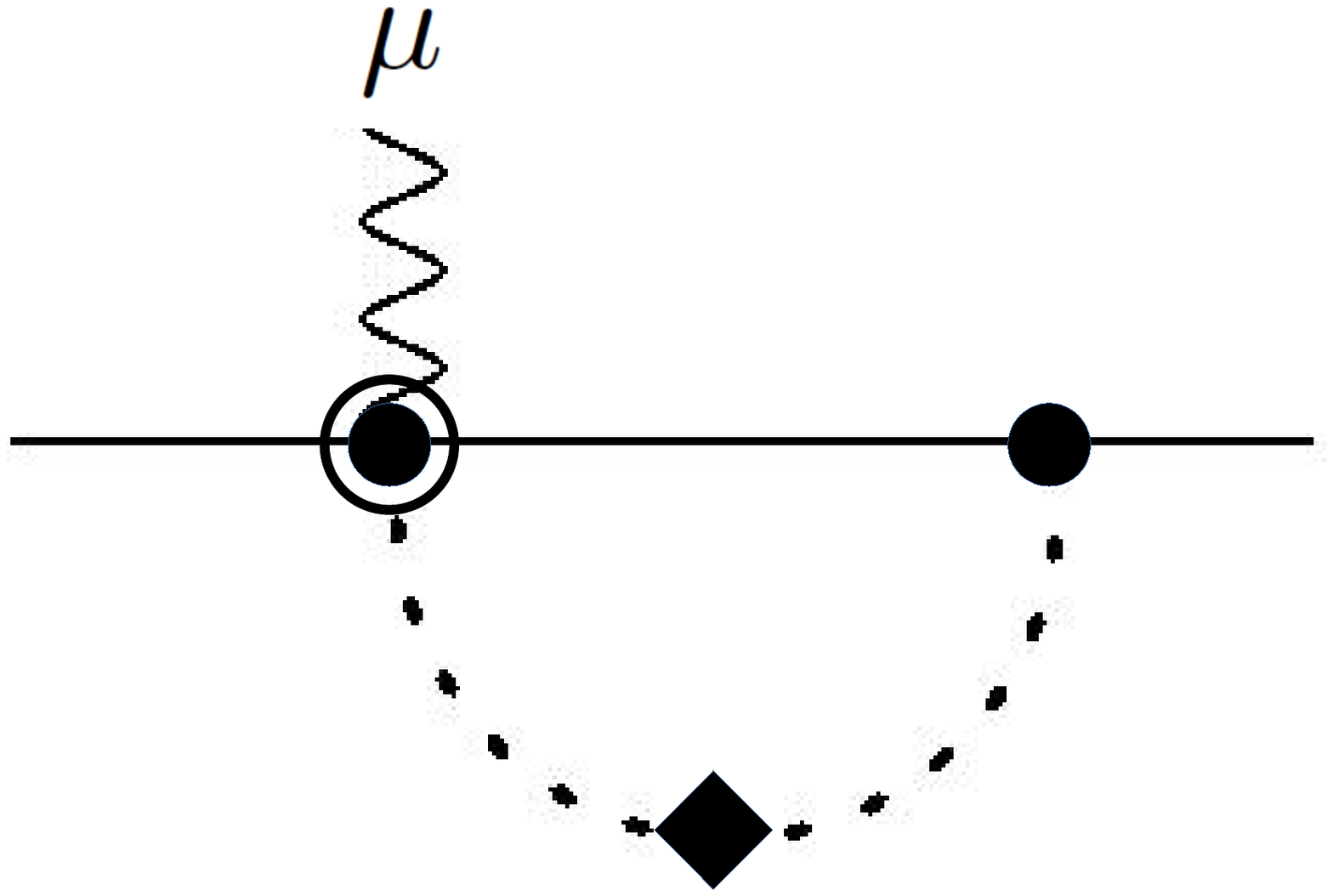}
		\caption{NLO diagrams for $\delta F_A^\mu$. Circled dot: Vertices from $\mathcal{L}_{\pi N}^{p^2}$. }
		\label{fig:NLO}
	\end{center}
\end{figure}
 
Following Ref.\cite{Cirigliano:2022hob}, we also compute the NLO corrections. The Feynman diagrams are given in Fig.\ref{fig:NLO} and the loop integrals are UV-finite. The resulting corrections are purely inner:
\begin{eqnarray}
(\delta F_A^\mu)_\text{NLO}&=&(\delta F_A^\mu)_\text{NLO}^\text{inner}\nonumber\\
&=&\frac{\alpha}{2\pi}\left[4\pi m_\pi Z_\pi\left(c_4-c_3+\frac{3}{8m_N}+\frac{9}{16m_N}\mathring{g}_A^2\right)-\frac{\pi m_\pi}{6m_N}(1+2\mathring{\mu}_p-2\mathring{\mu}_n)\right]F_A^\mu~.\nonumber\\
\end{eqnarray} 

\subsection{\label{sec:2pt3pt}Two- and three-point function}

Now we proceed to split the full $\delta F_A^\mu$ into two- and three-point function. One way to do it is to compute directly $\delta F_{A,\text{3pt}}^\mu$ from Feynman diagrams as described in Ref.\cite{Seng:2019lxf}; for that we need the Feynman rules of $\partial\cdot J_W$, on which equation of motion must be applied. In this paper we demonstrate an alternative approach, namely to compute $\delta F_{A,\text{2pt}}^\mu$ in EFT. We start from its definition in Eq.\eqref{eq:2pt}, and realize that the EFT only probes physics at $k\ll m_W$, so we take $m_W^2/(m_W^2-k^2)\rightarrow 1$ and adopt the dimensional regularization prescription for UV-divergences. This leads to the following EFT representation:
\begin{equation}
\delta F_{A,\text{2pt}}^\lambda=-e^2\mu^{4-d}\int\frac{d^dk}{(2\pi)^d}\frac{k^\lambda}{(k^2-m_\gamma^2)^2}(T^\mu_{\:\:\mu})^A~,
\end{equation}
where the superscript $A$ on $T^\mu_{\:\:\mu}$ denotes that only the axial component of the charged current is included.

\begin{figure}[t]
	\begin{center}
		\includegraphics[width=0.22\columnwidth]{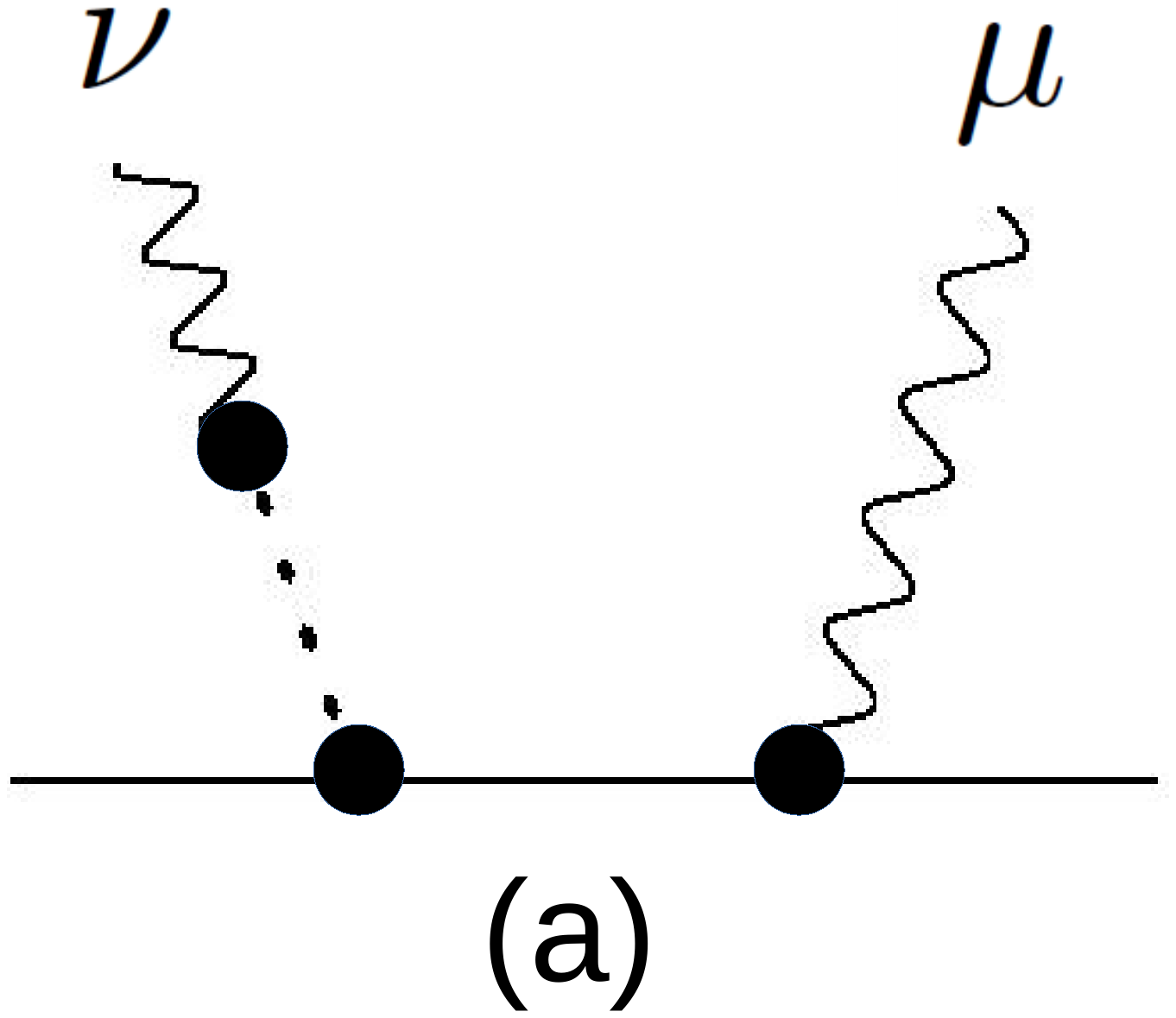}
		\includegraphics[width=0.22\columnwidth]{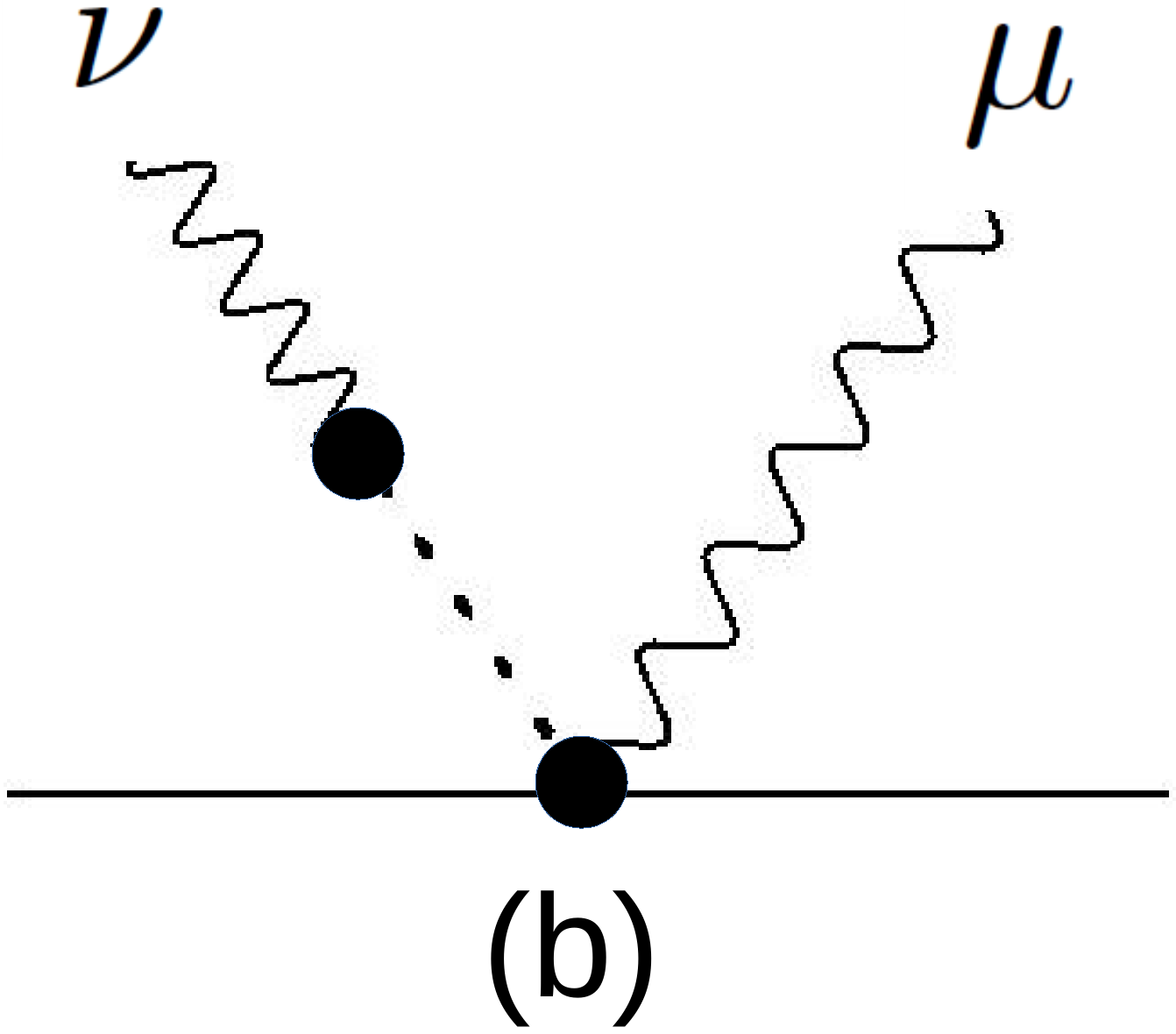}
		\includegraphics[width=0.22\columnwidth]{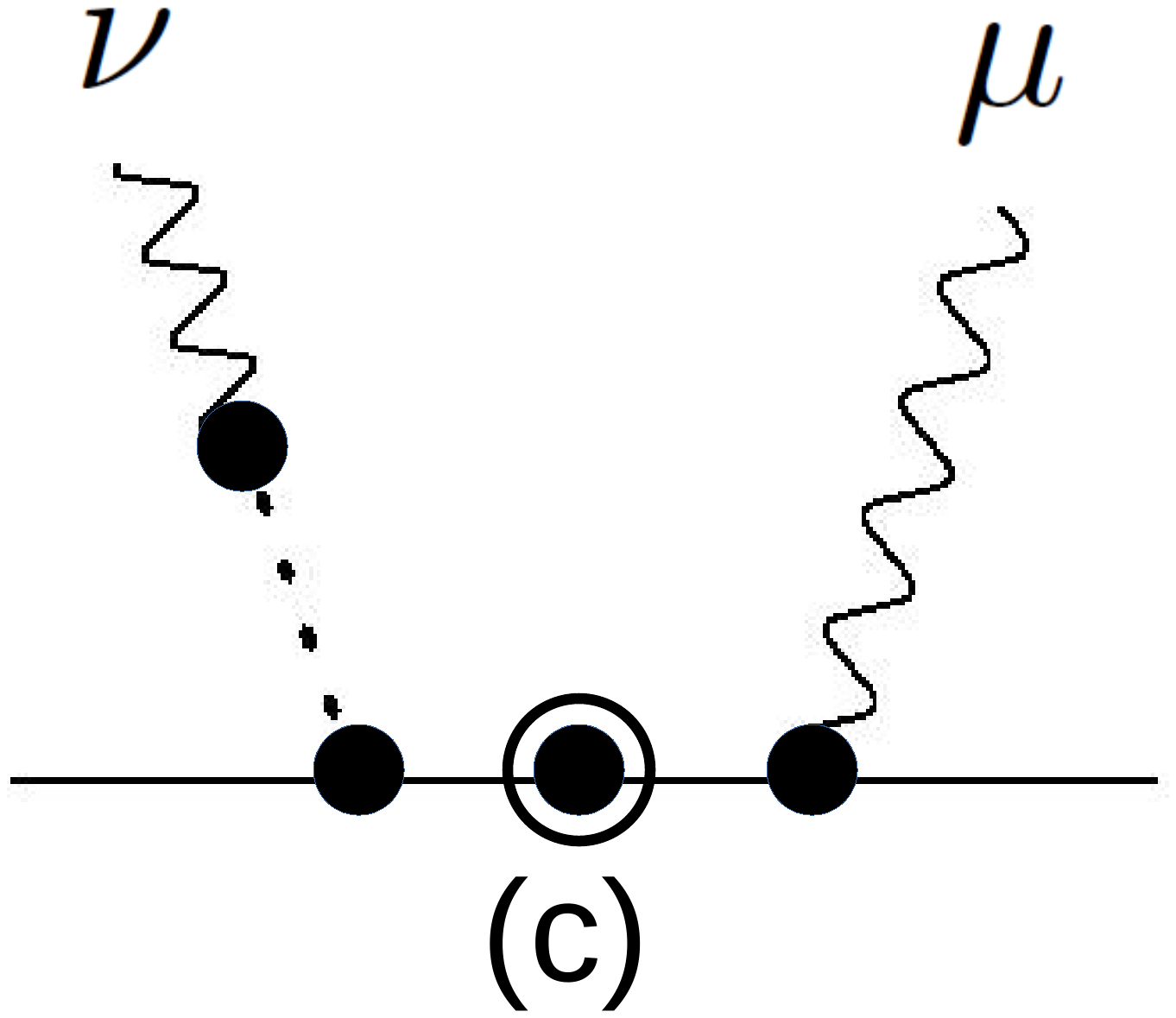}
		\includegraphics[width=0.22\columnwidth]{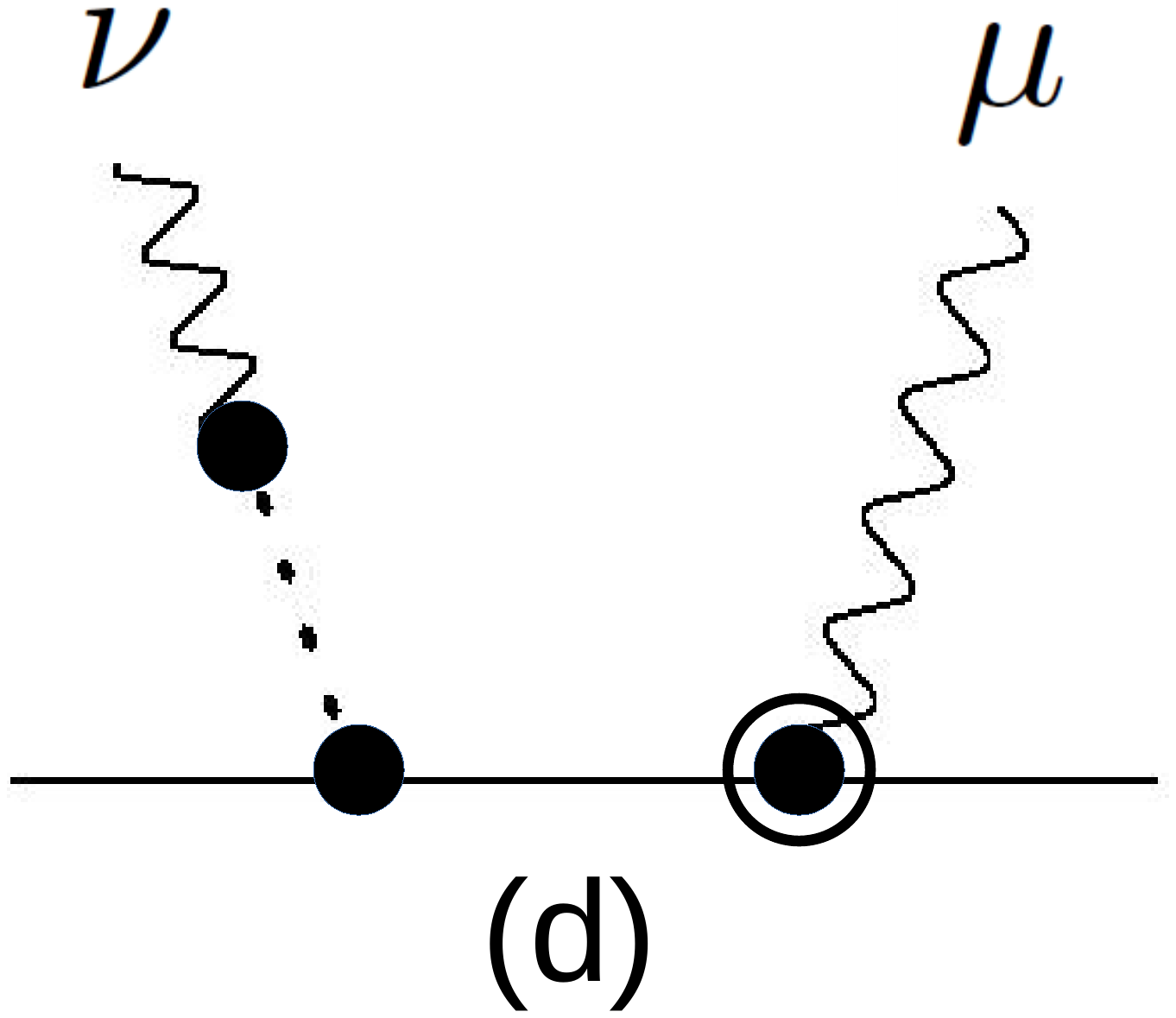}
		\caption{LO and NLO diagrams for $(T^{\mu\nu})^A$ with non-zero contribution to $\delta F_{A,\text{2pt}}^\mu$. }
		\label{fig:2pt}
	\end{center}
\end{figure}

According to its definition, it is obvious that the two-point function cannot depend on short-distance electromagnetic effect, i.e. those derived from $\mathcal{L}_\pi^{e^2}$ and $\mathcal{L}_{\pi N}^{e^2}$. Therefore, our task is to construct $(T^\mu_{\:\:\mu})^A$ at LO and NLO using the pure-QCD chiral Lagrangian. Non-vanishing contributions come from the Feynman diagrams in Fig.\ref{fig:2pt}, which give:
\begin{eqnarray}
(T^\mu_{\:\:\mu})_a^A&=&\frac{-2i\mathring{g}_A}{k^2-m_\pi^2}\left(1-\frac{\delta m_N}{v\cdot k}\right)\bar{u}_{p,v}S\cdot ku_{n,v}\nonumber\\
(T^\mu_{\:\:\mu})_b^A&=&\frac{2i\mathring{g}_A}{k^2-m_\pi^2}\bar{u}_{p,v}S\cdot k u_{n,v}\nonumber\\
(T^\mu_{\:\:\mu})_c^A&=&\frac{-i\mathring{g}_A}{k^2-m_\pi^2}\left(\frac{(v\cdot k)^2-k^2}{m_N v\cdot k}+\frac{2\delta m_N}{v\cdot k}\right)\bar{u}_{p,v}S\cdot k u_{n,v}\nonumber\\
(T^\mu_{\:\:\mu})_d^A&=&\frac{i\mathring{g}_A}{k^2-m_\pi^2}\frac{(v\cdot k)^2-k^2}{m_N v\cdot k}\bar{u}_{p,v}S\cdot k u_{n,v}~.\label{eq:Tmumu}
\end{eqnarray}
An interesting point to notice is that diagram (a), despite being LO, gives also an NLO contribution proportional to $\delta m_N$, namely the quark-mass correction to the nucleon mass from $\mathcal{L}_{\pi N}^{p^2}$, due to the on-shell condition of the external nucleon momenta. Here we assume the strong isospin symmetry: $\delta m_p=\delta m_n=\delta m_N$. 

We observe that the four terms in Eq.\eqref{eq:Tmumu} add up to zero, which implies that the full axial FF RC at both LO and NLO belongs to the three-point function. This is a major conclusion of this paper, which can also be verified independently by computing $\delta F_{A,\text{3pt}}^\mu$ using Feynman rules. After removing the outer corrections as in Eq.\eqref{eq:3ptsplit}, we obtain the desired EFT prediction of the inner correction to $g_A$ due to the three-point function:
\begin{equation}
\Delta_{R,\text{3pt}}^A=(\Delta_{R,\text{3pt}}^A)_\text{LO}+(\Delta_{R,\text{3pt}}^A)_\text{NLO}+...~,
\end{equation}
where
\begin{eqnarray}
(\Delta_{R,\text{3pt}}^A)_\text{LO}&=&\frac{\alpha}{2\pi}\left[Z_\pi(1+3\mathring{g}_A^2)\left(\ln\frac{\mu^2}{m_\pi^2}-1\right)-2\mathring{g}_A^2Z_\pi-\ln\frac{m_N^2}{m_\pi^2}+1\right.\nonumber\\
&&\left.-\frac{16\pi^2}{\mathring{g}_A}\left(g_1^r+g_2^r+\frac{g_{11}^r}{2}\right)\right]\nonumber\\
(\Delta_{R,\text{3pt}}^A)_\text{NLO}&=&\frac{\alpha}{2\pi}\left[8\pi m_\pi Z_\pi\left(c_4-c_3+\frac{3}{8m_N}+\frac{9}{16m_N}\mathring{g}_A^2\right)-\frac{\pi m_\pi}{3m_N}(1+2\mathring{\mu}_p-2\mathring{\mu}_n)\right]~,\nonumber\\\label{eq:DeltaRA}
\end{eqnarray}
which completes the missing piece in the SR + dispersion relation/lattice analysis. 

It is instructive to compare our result with Ref.\cite{Cirigliano:2022hob}. We find that, adding the expanded expression of $2\Box^A_\text{int}$ in Eq.\eqref{eq:Boxintexpand} to $\Delta_{R,\text{3pt}}^A$ reproduces exactly Eqs.(6), (7) in Ref.\cite{Cirigliano:2022hob}
(apart from differences in normalization) except the LECs $X_i$, $\tilde{X}_i$ and $g_9$ which correspond to the short-distance electroweak RC (i.e. $\Delta_R^U$ and $\Box_{\gamma W}^{V,A}$) as we will show later. 
Notice that, in this work we define the full $g_{V,A}$ in terms of the QED-corrected $n\rightarrow pe\bar{\nu}_e$ amplitude~\eqref{eq:Mren}, while in Ref.\cite{Cirigliano:2022hob} $g_{V,A}$ are defined as coupling constants in the pionless EFT Lagrangian $\mathcal{L}_{\slashed{\pi}}$, which will then be used to compute the decay amplitude coming from tree-level and photon-loop diagrams. As a consequence, $g_{V,A}$ in Ref.\cite{Cirigliano:2022hob} are scale-dependent but those here are not. Fortunately, what we really interested in is the ratio $\lambda=g_A/g_V$ which turns out to be the same in the two representations, because the photon-loop contribution to the inner RC in the pionless EFT is the same for $g_V$ and $g_A$, and thus cancels out in the ratio.

\section{\label{sec:LECs}Fixing the low energy constants}

An immediate application of the analysis above is to compare the predicted observables with those in the pure EFT representation. Since the two must be equivalent at $\mathcal{O}(\alpha)$, the matching allows us to pin down specific combinations of unknown LECs in the chiral Lagrangian. This idea was previously applied to semileptonic kaon decays, which successfully fixed several important LECs at $\mathcal{O}(e^2p)$ with $\sim 10$\% precision~\cite{Seng:2020jtz,Ma:2021azh}. 
 
The first observable to compare is the differential decay rate, i.e. Eq.\eqref{eq:differential} in this work and Eq.(26) in the Appendix of Ref.\cite{Cirigliano:2022hob}. Notice that, to compute the decay rate in an EFT one needs not only photons, but also leptons as dynamical DOFs. This introduces three types of (renormalized) LECs: $\{g_i^r\}$ from the pure hadron sector, $X_6^r$ from the pure lepton sector, and $\{\tilde{X}_i^r\}$ from the lepton-hadron sector; they enter the differential decay rate through the linear combination:
\begin{equation}
\hat{C}_V(\mu)\equiv 8\pi^2\left[-\frac{X_6^r}{2}+2\left(\tilde{X}_1^r-\tilde{X}_2^r\right)+g_9^r\right]~.
\end{equation} 
The comparison yields the following EFT representation of the famous ``nucleus-independent RC'':
\begin{equation}
\Delta_R^V=\frac{\alpha}{2\pi}\left[2\hat{C}_V(\mu)+\frac{5}{4}+3\ln\frac{\mu}{m_N}\right]~.
\end{equation} 
Substituting the expression of $\Delta_R^V$ in SR gives:
\begin{eqnarray}
	8\pi^{2}\left[-\frac{X_{6}^{r}}{2}+2(\tilde{X}_{1}^{r}-\tilde{X}_{2}^{r})+g_{9}^{r}\right]&=&\frac{3}{2}\ln\frac{m_Z}{\mu}+\frac{1}{2}\ln\frac{m_Z}{m_W}-\frac{5}{8}+\frac{1}{2}\tilde{a}_g+\frac{2\pi}{\alpha}\Box_{\gamma W}^V~.
\end{eqnarray}
Notice that we do not include $\delta_\text{HO}^\text{QED}$ in the matching because it was not a part of the fixed-order HBChPT calculation. 
For the vector box diagram $\Box_{\gamma W}^V$, we can either take $3.85(9)\times 10^{-3}$ obtained from an average over several non-lattice determinations~\cite{Gorchtein:2023srs}, or $3.65(8)\times 10^{-3}$ from a direct lattice calculation~\cite{Ma:2023kfr}, the two exhibit a $1.7\sigma$ deviation.  
Here we decide to average them: $\Box_{\gamma W}^V=3.74(10)\times 10^{-3}$, where the uncertainty is enlarged by a scale factor $S=1.7$ following the usual practice by the Particle Data Group~\cite{ParticleDataGroup:2022pth}.
We then obtain the numerical value of the following LEC combination at $\mu=m_N$:
\begin{equation}
	8\pi^{2}\left[-\frac{X_{6}^{r}}{2}+2(\tilde{X}_{1}^{r}-\tilde{X}_{2}^{r})+g_{9}^{r}\right](\mu=m_{N})=9.48(13)_\text{HO}(9)_\text{box}~,
\end{equation}
where the uncertainties come from terms of higher chiral orders (estimated by multiplying the central value by $m_\pi^2/\Lambda_\chi^2$, with $\Lambda_\chi=4\pi F_0$ the chiral symmetry breaking scale) and $\Box_{\gamma W}^V$ respectively.

The second observable to compare is the axial-to-vector ratio $\lambda$, i.e. Eq.\eqref{eq:ratio} of this work and Eq.(12) in Ref.\cite{Cirigliano:2022hob}. Notice that $g_A$ in EFT depends on a separate linear combination of LECs:
\begin{equation}
\hat{C}_A(\mu)\equiv 8\pi^2\left[-\frac{X_6^r}{2}-\frac{1}{\mathring{g}_A}\left[\tilde{X}_3^r+\left(g_1^r+g_2^r+\frac{g_{11}^r}{2}\right)\right]\right]~.
\end{equation} 
The comparison yields the following matching of LECs:
\begin{eqnarray}
8\pi^2\left[-\frac{X_6^r}{2}-\frac{1}{\mathring{g}_A}\tilde{X}_3^r\right]&=&\frac{3}{2}\ln\frac{m_Z}{\mu}+\frac{1}{2}\ln\frac{m_Z}{m_W}-\frac{5}{8}+\frac{1}{2}\tilde{a}_g+\frac{2\pi}{\alpha}\Box_{\gamma W}^A~,
\end{eqnarray}
where we have made use of the expanded form of $\Box^A_\text{int}$ in Eq.\eqref{eq:Boxintexpand}.
The dispersion relation~\cite{Gorchtein:2021fce} and lattice~\cite{Ma:2023kfr} determination of the axial box $\Box_{\gamma W}^A$ return $\Box_{\gamma W}^A$: $3.96(6)\times 10^{-3}$ and $3.72(14)\times 10^{-3}$ respectively, showing a $1.6\sigma$ discrepancy. Averaging the two yields $\Box_{\gamma W}^A=3.92(9)\times 10^{-3}$ where the uncertainty is enlarged by a scale factor $S=1.6$. It returns:
\begin{equation}
	8\pi^{2}\left[-\frac{X_{6}^{r}}{2}-\frac{1}{\mathring{g}_{A}}\tilde{X}_{3}^{r}\right](\mu=m_{N})=9.64(13)_\text{HO}(8)_\text{box}~.
\end{equation}

Up to this point, the only undetermined combination of LECs relevant to neutron beta decay is $g_1^r+g_2^r+g_{11}^r/2$ that appears in the three-point function. In principle, since $\delta F_\text{3pt}^\mu$ is rigorously defined in terms of single-nucleon matrix elements in Eq.\eqref{eq:3ptME}, one should be able to compute it with lattice QCD; but in practice, what we call ``three-point function'' here is in fact a ``five-point function'' in lattice QCD (two external states + three current insertions), which is extremely difficult to handle. An alternative approach is to compute the full axial current matrix element $\langle p|(J_W^\mu)_A|n\rangle$ in the presence of electromagnetic interactions, which gives us directly $\delta F_A^\mu$. This is possible using, e.g. lattice QCD + QED with massive photons~\cite{Endres:2015gda}, which is also referred to as QED$_\text{M}$. We have shown in Sec.\ref{sec:2pt3pt} that the relation $\delta F_A^\mu=\delta F_{A,\text{3pt}}^\mu$ holds at least up to NLO; so, a lattice calculation of $\delta F_A^\mu$ provides a first-principles determination of $\Delta_{R,\text{3pt}}^A$ after removing the IR-divergent outer corrections:
\begin{equation}
	\langle p|(J_W^\mu)_A|n\rangle_{\text{QCD}+\text{QED}_\text{M}}=\left[1-\frac{\alpha}{4\pi}\left(\ln\frac{m_\gamma^2}{m_N^2}+2\right)+\frac{1}{2}\Delta_{R,\text{3pt}}^A\right]\langle p|(J_W^\mu)_A|n\rangle_{\text{QCD}}~.\label{eq:QEDM}
\end{equation}
In fact, \eqref{eq:ratio} and \eqref{eq:QEDM} are the only equations needed as far as the RC to $\lambda$ is concerned. Substituting the lattice-deduced value of $\Delta_{R,\text{3pt}}^A$ to Eq.\eqref{eq:DeltaRA} yields the numerical value of the combination $g_1^r+g_2^r+g_{11}^r/2$ which can be used to analyze other processes. 

We conclude this section by stressing that SR provides a useful basis for lattice studies of RC in beta decays because lepton fields are decoupled from its master formula, thus bypassing the difficulty to introduce leptonic DOFs on lattice. To give a recent example, the RC in semileptonic kaon decays ($K_{\ell 3}$) are determined to $10^{-4}$ precision under this framework~\cite{Seng:2021boy,Seng:2021wcf,Seng:2022wcw} where the only required lattice inputs are the $K\pi$ and $\pi\pi$ box diagrams that are relatively straightforward to compute on lattice~\cite{Feng:2020zdc,Ma:2021azh}.
In contrast, a full-fledged lattice calculation of the $K_{\ell 3}$ RC (virtual and real) with dynamical photons and leptons has a projected time scale of $\sim$ 10 years to reach a $10^{-3}$ precision~\cite{BoyleSnowmass}. We expect the same advantage to also hold in the nucleon sector.

\section{\label{sec:summary}Summary}

Recent years have witnessed tremendous improvements in the experimental determination of the neutron lifetime $\tau_n$ and the axial-to-vector ratio $\lambda$, both reaching a precision level of $10^{-4}$. To make full use of these results for precision tests of SM, a control of the SM RC uncertainty at $10^{-4}$ is necessary. In early days, the RC to the spin-independent part of the neutron beta decay received more attention as it affects the extraction of $V_{ud}$. Recently, due to breakthroughs in first-principles calculation of $\mathring{g}_A$ that approaches sub-percent level precision, the understanding of the spin-dependent part of the RC becomes equally important. Recent EFT analysis unveiled a percent-level correction to $g_A$ which was not covered by the previous dispersive analysis of $\gamma W$-box diagrams. It plays a crucial role when the lattice-calculated $\mathring{g}_A$ is compared to experimentally-measured $\lambda$ for the search of new physics.

In principle, the full RC to inclusive neutron decay can be studied on lattice by including photonic and leptonic DOFs, which is extremely challenging in practice. It is easier to include only photonic DOFs, and such a setup is ideal to compute the QED-corrected single-nucleon FFs, in which the aforementioned large RC resides. Correspondingly, one needs an EFT description for just the RC to the axial FF instead of the full RC, which is attempted in this paper. Furthermore, we merge this EFT calculation with the well-developed current algebra formalism of RC, which is extremely powerful in controlling all the remaining one-loop RC when combine with dispersion relation analysis. By comparing the expressions between our hybrid analysis and the full EFT calculation, we are able to determine numerically a subset of LECs that correspond to the vector and axial $\gamma W$-box diagrams. Furthermore, we show that a future lattice calculation of $\delta F_A^\mu$ will provide the last missing piece $\Delta_{R,\text{3pt}}^A$ in the hybrid method, namely the inner correction to $g_A$ from the three-point function. 

Our analysis provides an elegant framework for non-perturbative studies of the neutron decay RC, where only a minimal amount of lattice QCD inputs are required. It will be useful for current and future tests of SM and the search of BSM physics at the precision frontier.

\begin{acknowledgments}
	
We thank Vincenzo Cirigliano, Wouter Dekens, Emanuele Mereghetti and Oleksandr Tomalak for many useful discussions, especially for pointing out the missing term $\Box^A_\text{int}$ in the first arXiv version. 
The work of C.-Y.S. is supported in
part by the U.S. Department of Energy (DOE), Office of Science, Office of Nuclear Physics, under the FRIB Theory Alliance award DE-SC0013617, and by the DOE grant DE-FG02-97ER41014. We acknowledge support from the DOE Topical Collaboration ``Nuclear Theory for New Physics'', award No. DE-SC0023663. 
  
\end{acknowledgments}

\bibliography{gA_ref}

\end{document}